\documentclass[12pt]{article}
\usepackage{a4wide,epsfig,psfrag,amsmath,amssymb,cite,scalefnt}
\usepackage{color}

\newcommand{\gsim}{\;\rlap{\lower 3.5 pt \hbox{$\mathchar \sim$}} \raise 1pt
 \hbox {$>$}\;}
\newcommand{\lsim}{\;\rlap{\lower 3.5 pt \hbox{$\mathchar \sim$}} \raise 1pt
 \hbox {$<$}\;}

\newcommand{\mufs}{\mu_F}

\begin{document}

\title{\vskip-3cm{\baselineskip14pt
    \begin{flushleft}
      \normalsize SFB/CPP-09-116\\
      \normalsize TTP09-43
  \end{flushleft}}
  \vskip1.5cm
  Finite top quark mass effects in NNLO Higgs boson production at LHC
}

\author{\small 
  Alexey Pak,
  Mikhail Rogal,
  Matthias Steinhauser
  \\
  {\small\it Institut f{\"u}r Theoretische Teilchenphysik}\\
  {\small\it Karlsruhe Institute of Technology (KIT)}\\
  {\small\it 76128 Karlsruhe, Germany}
}

\date{}

\maketitle

\thispagestyle{empty}

\begin{abstract}
We present next-to-next-to-leading order corrections to the inclusive
production of the Higgs bosons at the CERN Large Hadron Collider (LHC)
including finite top 
quark mass effects. Expanding our analytic results for the partonic
cross section around the soft limit we find agreement with a very 
recent publication by Harlander and Ozeren~\cite{Harlander:2009mq}.

\medskip


\end{abstract}

\thispagestyle{empty}


\newpage


\section{Introduction}

With the launch of the Large Hadron Collider (LHC) at CERN the particle
physics experiments enter a new energy domain with the hope to discover
new phenomena. Experimental observations are expected to provide
hints for the still open questions within the Standard Model.
Among these questions is the mechanism of the electroweak symmetry
breaking which in most practical theories provides masses to the
particles. The traditional implementation of the symmetry breaking
implies the existence of a new particle, the Higgs boson, which so far
has not been detected in high-energy collisions.

For the intermediate Higgs boson mass, favoured by the results
of indirect searches, the most important production channel at a
hadron collider is gluon fusion, $gg\to H$, mediated by a top quark loop.
During the last 20 years enormous efforts have been made to evaluate
higher order corrections to this process.

The leading order (LO) result has been presented in
Refs.~\cite{Wilczek:1977zn,Ellis:1979jy,Georgi:1977gs,Rizzo:1979mf}
and already almost 15 years ago also the next-to-leading order (NLO)
QCD corrections became available~\cite{Dawson:1990zj,Spira:1995rr}.
More recently also the next-to-next-to-leading order (NNLO) corrections
have been
evaluated~\cite{Harlander:2000mg,Harlander:2002wh,Anastasiou:2002yz,Ravindran:2003um}.
While the NLO results are exact in the top quark and Higgs boson masses
(in this context, see also Ref.~\cite{Harlander:2005rq}),
the NNLO results rely on the effective theory built in the 
limit of the large top quark mass (see, e.g.,
Refs.~\cite{Chetyrkin:1997un,Steinhauser:2002rq}
for the three-loop corrections to the effective $ggH$ coupling).
It is well known that this approximation works surprisingly well at NLO,
leading to deviations from the exact
result that are less than 2\% for $M_H<2M_t$~\cite{Harlander:2003xy}.
It is one of the aims of the present paper to investigate the validity of the
large top quark mass approximation at NNLO.

During the last few years there appeared several improvements over the 
fixed-order calculation. Among them is the soft-gluon resummation to
next-to-next-to-leading~\cite{Catani:2003zt} and
next-to-next-to-next-to-leading~\cite{Moch:2005ky}
logarithmic orders and the identification (and resummation) of certain
$\pi^2$ terms~\cite{Ahrens:2008nc} which significantly improves the
perturbative series.
We furthermore want to mention Ref.~\cite{Marzani:2008az} where the
gluon-gluon channel has been considered in the limit of large
center-of-mass energy $\sqrt{\hat{s}}$.
Recent numerical predictions of Higgs boson production in gluon fusion
both at the Tevatron and the LHC are summarized in Ref.~\cite{deFlorian:2009hc}.

Very recently the effects of the finite top quark mass on the Higgs boson
production in hadron colliders have been reported in
Ref.~\cite{Harlander:2009mq}. The authors of that reference
carried out an asymptotic expansion of the corresponding production diagrams
and obtained results for the first four terms in the $\rho = M_H^2/M_t^2$ expansion,
where the secondary expansion around the soft limit (i.e. for
$x = M_H^2/\hat{s} \to 1$) has been performed to a sufficiently high order.

In this paper we present the results of an independent calculation of these
finite top mass effects, confirming the conclusions of
Ref.~\cite{Harlander:2009mq}. 
Similarly to Ref.~\cite{Harlander:2009mq}, we apply asymptotic expansion
in $1/M_t$ to the full QCD diagrams and evaluate a few first terms of the 
series. However, our results are not expanded near the soft limit and the
$x$-dependence of the cross section (which is valid below the top pair
threshold, as discussed further) is retained.

The remainder of the paper is organized as follows. In the next
Section we discuss the cross sections in the individual channels
of partonic reactions at NLO and NNLO. We consider in particular the 
behaviour near $x\to 0$ and describe a method to 
extrapolate our results to the limit of large $\hat{s}$.
In Section~\ref{sec::hadr} we use the results of
Section~\ref{sec::part} and numerically evaluate the hadronic cross
section for the LHC and the Tevatron.
Our conclusions are presented in Section~\ref{sec::concl}.


\section{\label{sec::part}Partonic cross section}


\subsection{Notations and calculation details}

We introduce the following notation for the partonic cross section:
\begin{eqnarray}
  \hat{\sigma}_{ij\to H + X} &=& \hat{A}_{\rm LO} \left(
    \Delta_{ij}^{(0)} + \frac{\alpha_s}{\pi}~ \Delta_{ij}^{(1)}  
    + \left(\frac{\alpha_s}{\pi}\right)^2 \Delta_{ij}^{(2)} + \ldots
  \right)
  \,,
  \label{eq::hatsigma}
\end{eqnarray}
with 
\begin{eqnarray}
  \hat{A}_{\rm LO} &=& \frac{G_F~\alpha_s^2}{288\sqrt{2}\pi} 
  f_0(\rho,0)
\end{eqnarray}
and $ij$ denoting one of the possible initial states:
$gg$, $qg$, $q\bar{q}$, $qq$, or $qq^\prime$, where $q$ and $q^\prime$
stand for (different) massless quark flavours.\footnote{It is understood
that ghosts are always considered together with gluons.} At NNLO the Higgs boson 
in the final state may be accompanied by zero, one or two gluons, or by 
a light quark pair. In general, the quantities $\Delta^{(k)}_{ij}$ depend on 
$x$ and $\rho$. Leading order mass dependence 
is then described by the function $f_0(\rho,0)$ given in Eq.~(4)
of Ref.~\cite{Pak:2009bx}.

Factoring out the exact LO top quark mass dependence as in
Eq.~(\ref{eq::hatsigma}) is a common practice. In what follows, by 
the ``infinite top quark mass approximation'' we mean that only the quantities
$\Delta_{ij}^{(k)}$ are evaluated for $M_t\to \infty$, but $\hat{A}_{\rm LO}$
remains exact in $M_t$.

At the LO only $\Delta_{gg}^{(0)}$ is different from zero and given by
\begin{eqnarray}
  \Delta_{gg}^{(0)} &=& \delta(1-x).
\end{eqnarray}

At the NLO the functions $\Delta^{(1)}_{gg}$, $\Delta^{(1)}_{qg}$ and
$\Delta^{(1)}_{q\bar{q}}$ are not zero, and at the NNLO one has to consider 
all five contributions: $\Delta^{(2)}_{gg}$, $\Delta^{(2)}_{qg}$,
$\Delta^{(2)}_{q\bar{q}}$, $\Delta^{(2)}_{qq}$, and $\Delta^{(2)}_{qq^\prime}$.

The first results on the $\rho$ dependence of NNLO cross sections appeared in 
Refs.~\cite{Pak:2009bx} and~\cite{Harlander:2009bw}, where the virtual part 
of $\Delta^{(2)}_{gg}$ was evaluated to $\mathcal{O}(\rho^4)$ and $\mathcal{O}(\rho^2)$,
respectively. In this paper we adhere to the notations of Ref.~\cite{Pak:2009bx}; 
in particular, we use $\alpha_s^{(5)}$ and the on-shell top quark mass.

\begin{figure}[t]
  \centering
  \includegraphics[width=0.3\linewidth]{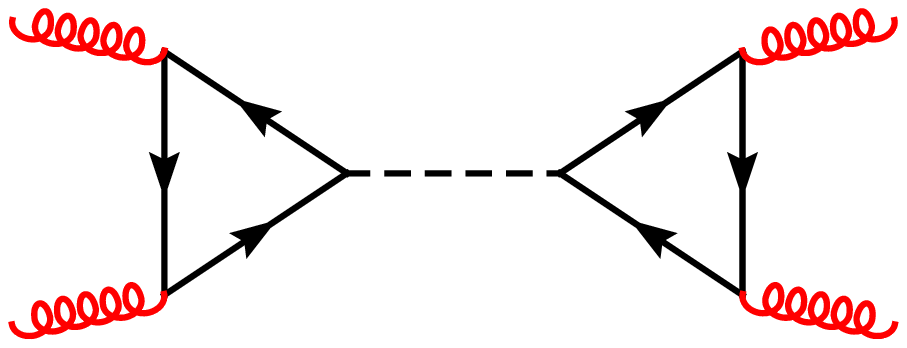}\hfill
  \includegraphics[width=0.3\linewidth]{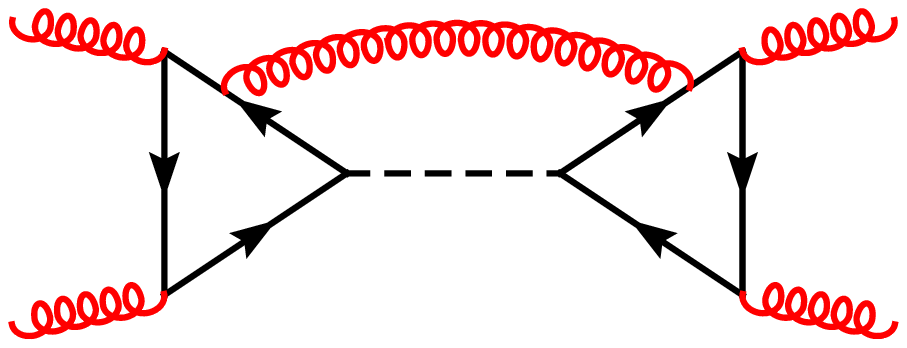}\hfill
  \includegraphics[width=0.3\linewidth]{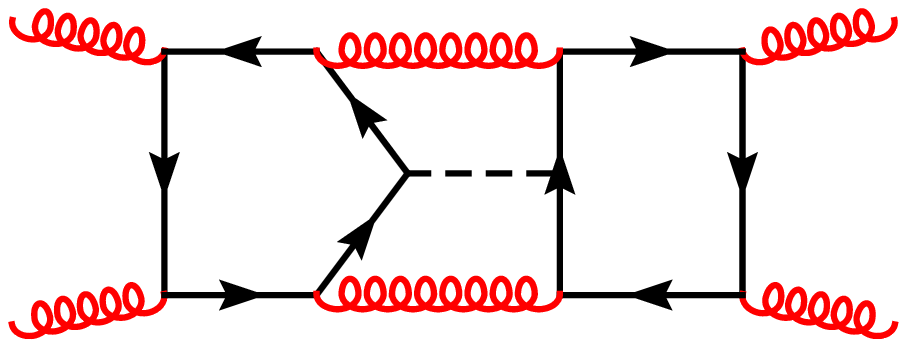}
  \\[1em]
  \includegraphics[width=0.3\linewidth]{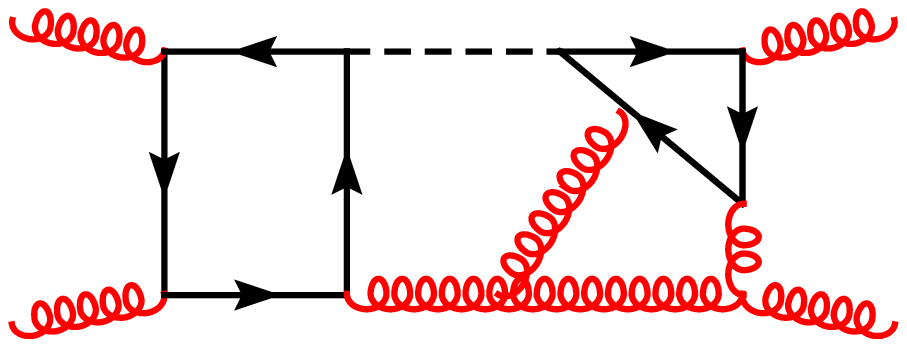}\hfill
  \includegraphics[width=0.3\linewidth]{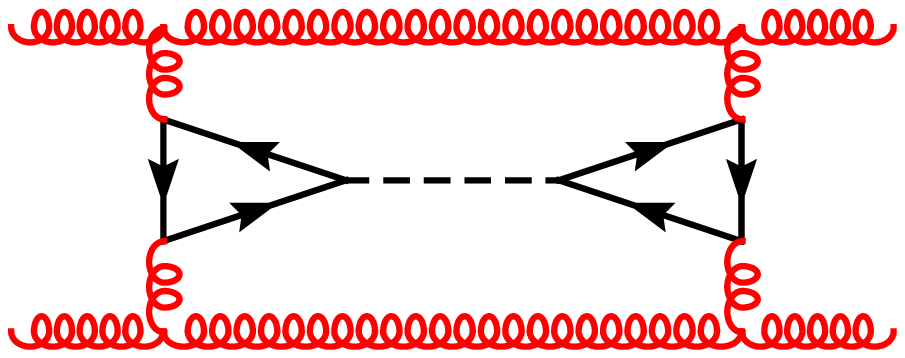}\hfill
  \includegraphics[width=0.3\linewidth]{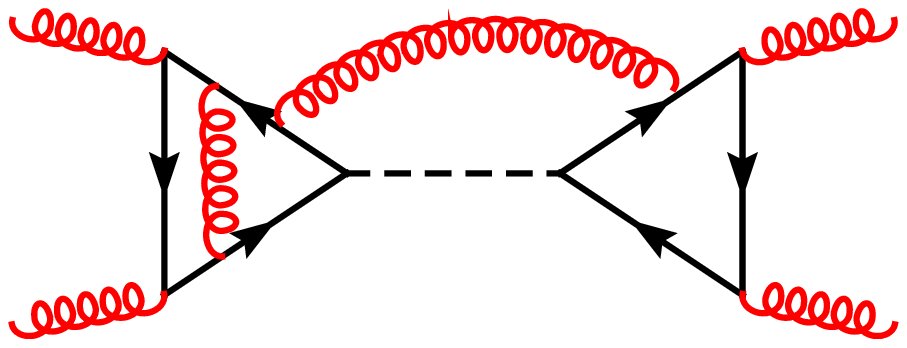}
  \caption[]{\label{fig::diag}Sample forward scattering diagrams
    whose cuts correspond to the LO, NLO and NNLO corrections to
    $gg\to H$. Dashed, curly and solid lines represent Higgs bosons,
    gluons and top quarks, respectively.}
\end{figure}

Since at the LO only the virtual contribution is present, we do not discuss it further.
In order to evaluate the real corrections to $\Delta_{ij}^{(k)}$, we exploit the
optical theorem and compute the imaginary part of the four-point 
amplitudes $ij\to ij$. At NLO and NNLO this requires the evaluation of
three- and four-loop diagrams, respectively. Some sample diagrams are
shown in Fig.~\ref{fig::diag}. Note that only the cuts dissecting the Higgs
boson propagator and one or two massless lines need to be included. 

We generate the diagrams with the help of {\tt QGRAF}~\cite{Nogueira:1991ex}
supplemented by additional scripts that eliminate the vanishing graphs.
At the next step we apply the asymptotic expansion (see,
e.g., Ref.~\cite{Smirnov:2002pj}) in the limit $M_t^2\gg
\hat{s},M_H^2$, implemented in two independent programs:
{\tt q2e}/{\tt exp}~\cite{Harlander:1997zb,Seidensticker:1999bb},
and an in-house Perl implementation.

This procedure factorizes the original triple-scale forward scattering functions
into massive vacuum integrals (with a single scale $M_t$) up to three loops, that can be
evaluated with {\tt MATAD}~\cite{Steinhauser:2000ry}, and four-point one- and two-loop 
integrals dependent on $\hat{s}$ and $M_H$. After the reduction performed with
the Laporta
algorithm~\cite{Laporta:1996mq,Laporta:2001dd} we arrive to a set of master
integrals that have been studied in Ref.~\cite{Anastasiou:2002yz} (Appendix~B).
Unfortunately, that reference contains a number of misprints. We independently 
evaluated these integrals by a combination of soft expansion\footnote{We
  acknowledge help with cross checks of the soft expansion by Robert
  Harlander and Kemal Ozeren.} and differential 
equation methods (a detailed comparison to be provided elsewhere).

Using these techniques we have been able to compute NNLO contributions 
through $\mathcal{O}(\rho^2)$ for the channels $gg$ and $qg$, and through
$\mathcal{O}(\rho^3)$ for $q\bar{q}$, $qq$ and $qq^\prime$. 


\subsection{NLO results}

The integral representation of the NLO corrections to the partonic cross
section can be found in Refs.~\cite{Dawson:1990zj,Spira:1995rr}.
In analytic form the $\rho^0$ and $\rho^1$ terms ($gg$ channel) were presented
in Refs.~\cite{Spira:1995rr} and~\cite{Dawson:1993qf}, respectively. In
Ref.~\cite{Harlander:2009mq} terms up to ${\cal O}(\rho^3)$ have been
demonstrated. Here it is convenient for us to include terms through 
$\mathcal{O}(\rho^4)$ given by
\begin{eqnarray}
  \Delta_{gg}^{(1)} &=&
          - {11 \over 2} (1-x)^3
          + 6~ {\rm H}(0;x) \left(2 x - x^2 + x^3 - {1 \over 1-x}\right)
          + 12~ {\rm H}(1;x) \left(2 x - x^2 + x^3\right)
  \nonumber \\ &+&
            \delta(1-x) \left({11 \over 2} + \pi^2\right)
          + 12 \left\lfloor {\ln(1-x) \over 1-x} \right\rfloor_+
  + \rho \left(
          - {3 x \over 20}
          + {3 x^2 \over 20}
          + {34 \over 135} \delta(1-x)
  \right)
  \nonumber \\ &+&
    \rho^2 \left(
            {37 \over 11200 x^2}
          + {39 \over 11200 x}
          - {159 \over 11200}
          + {151 x \over 11200}
          - {17 x^2 \over 2800}
          + {3553 \over 113400} \delta(1-x)
  \right)
  \nonumber \\ &+&
    \rho^3 \left(
            {23 \over 12000 x^3}
          - {41 \over 10500 x^2}
          + {1163 \over 288000 x}
          - {2267 \over 672000}
          + {1073 x \over 336000}
          - {377 x^2 \over 201600}
  \right. \nonumber \\ &+& \left.
            {917641 \over 190512000} \delta(1-x)
  \right)
    + \rho^4 \left(
            { 3347 \over 77616000  x^4}
          + { 72587 \over 141120000 x^3}
  \right. \nonumber \\ &-& \left.
            { 16912657 \over 12418560000 x^2}
          + { 58831963 \over 37255680000 x}
          - { 2097223 \over 1774080000}
          + { 10057661 x \over 12418560000}
  \right. \nonumber \\ &-& \left.
            { 749741 x^2 \over 1862784000}
          + { 208588843 \over 251475840000} \delta(1-x)
  \right)
  \,,
  \label{eqn:nlogg}
\end{eqnarray}
\begin{eqnarray}
  \Delta_{qg}^{(1)} &=&
          - 1 + 2 x - {x^2 \over 3}
          - {4 - 4 x + 2 x^2 \over 3} \left({\rm H}(0;x) + 2 {\rm H}(1;x)\right)
  \nonumber \\ &+& \rho \left(
          - {22 \over 135 x}
          + {11 \over 45}
          - {11 x \over 90}
          + {11 x^2 \over 270}
  \right)
  \nonumber \\ &+&
    \rho^2 \left(
            {3487 \over 259200 x^2}
          - {481 \over 12600 x}
          + {5171 \over 151200}
          - {859 x \over 64800}
          + {457 x^2 \over 120960}
  \right)
  \nonumber \\ &+&
    \rho^3 \left(
          - {539 \over 324000 x^3}
          + {98591 \over 15552000 x^2}
          - {15751 \over 1701000 x}
          + {54857 \over 9072000}
          - {5357 x \over 2721600}
  \right. \nonumber \\ &+& \left.
            {7861 x^2 \over 15552000}
  \right)
    + \rho^4 \left(
            { 107369 \over 436590000  x^4}
          - { 80543 \over  68040000 x^3}
          + { 219381011\over 95800320000  x^2}
  \right. \nonumber \\ &-& \left.
            { 286017499\over 125737920000 x}
          + { 717887 \over 609638400}
          - { 889451x \over  2661120000}
          + { 159415681x^2 \over 2011806720000 }
  \right)
  \,,
  \label{eqn:nloqg}
\end{eqnarray}
\begin{eqnarray}
  \Delta_{q\bar{q}}^{(1)} &=&
            {32 \over 27} (1-x)^3
  + \rho \left(
            {88 \over 405 x}
          - {88 \over 135}
          + {88 x \over 135}
          - {88 x^2 \over 405}
  \right)
  \nonumber \\ &+&
    \rho^2 \left(
            {3487 \over 85050 x^2}
          - {9619 \over 85050 x}
          + {529 \over 5670}
          - {961 x \over 85050}
          - {421 x^2 \over 42525}
  \right)
  \nonumber \\ &+&
    \rho^3 \left(
            {49 \over 6075 x^3}
          - {107209 \over 5103000 x^2}
          + {2006 \over 127575 x}
          - {1511 \over 850500}
          + {8 x \over 91125}
          - {5573 x^2\over 5103000}
  \right)
  \nonumber \\ &+&
    \rho^4 \left(
            {107369 \over 65488500 x^4}
          - {40252 \over 9823275 x^3}
          + {912689 \over 317520000 x^2}
          - {2390203 \over 8573040000 x}
  \right. \nonumber \\ &-& \left.
            {837833 \over 31434480000}
          + {270467 x \over 6286896000}
          - {7256033 x^2\over 47151720000}
  \right)
  \,.
  \label{eqn:nloqq}
\end{eqnarray}

\begin{figure}[t]
  \centering
  \begin{tabular}{cc}
    \includegraphics[width=0.5\linewidth]{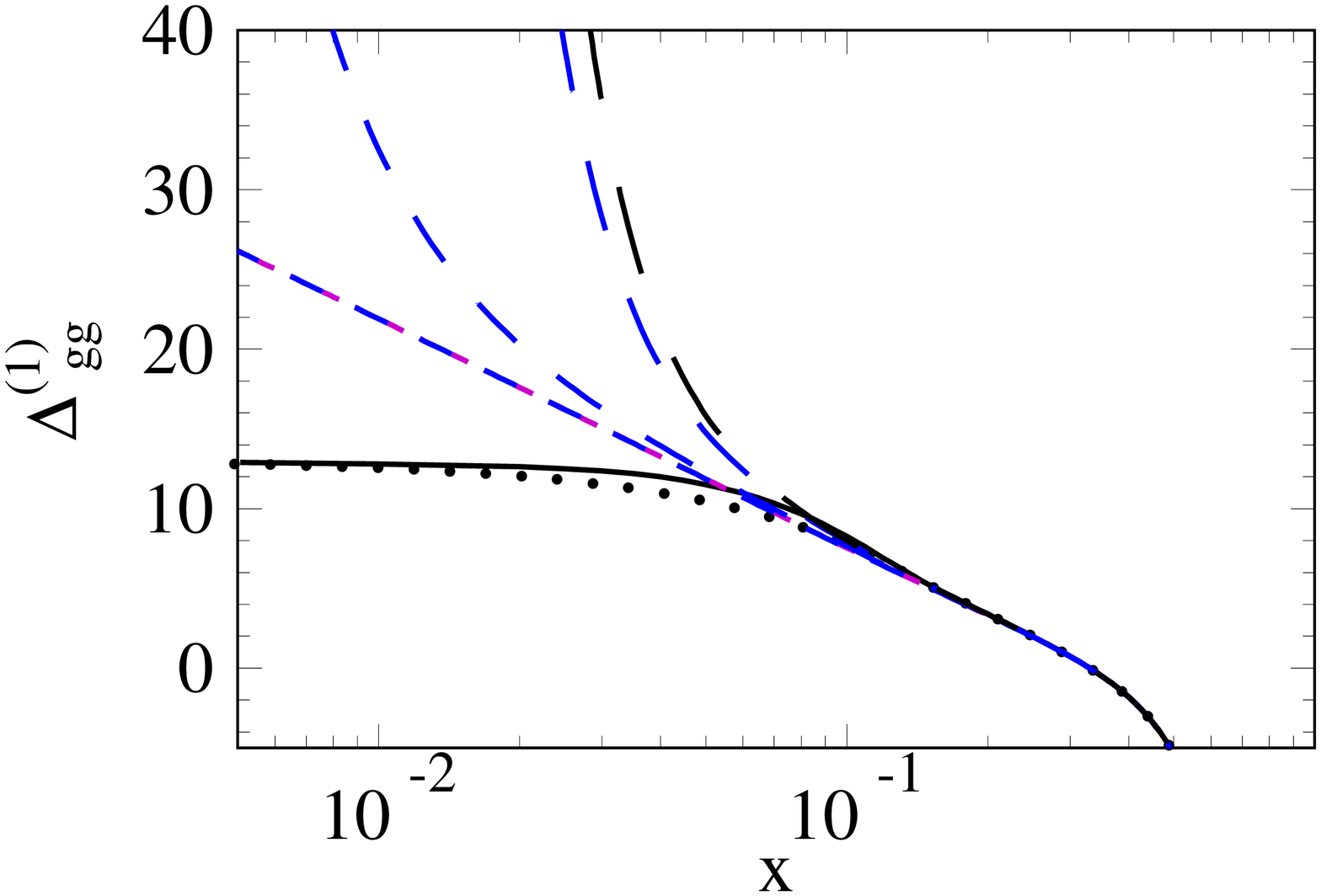}
    &
    \includegraphics[width=0.5\linewidth]{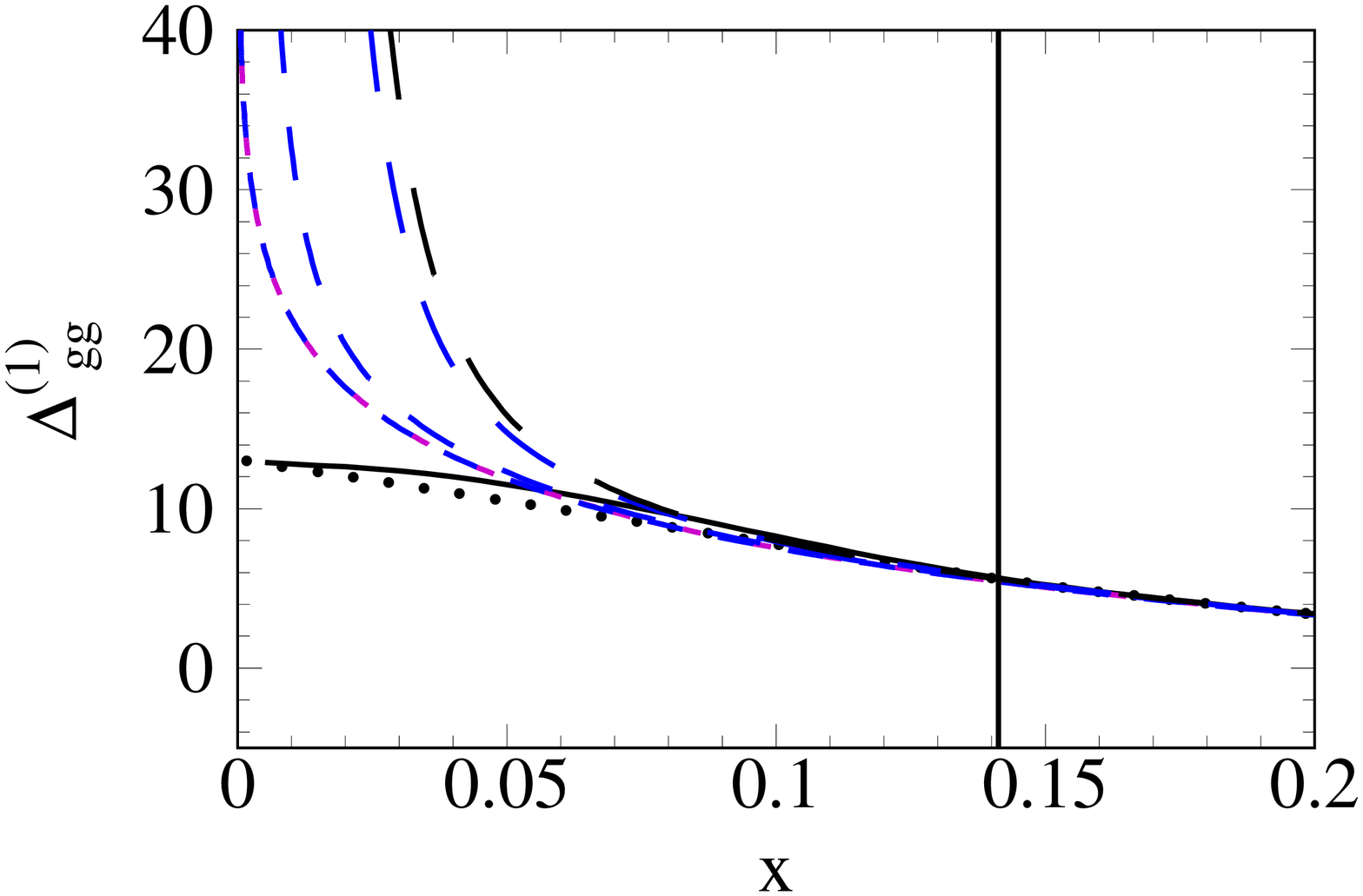}
    \\ (a) & (b) \\
    \includegraphics[width=0.5\linewidth]{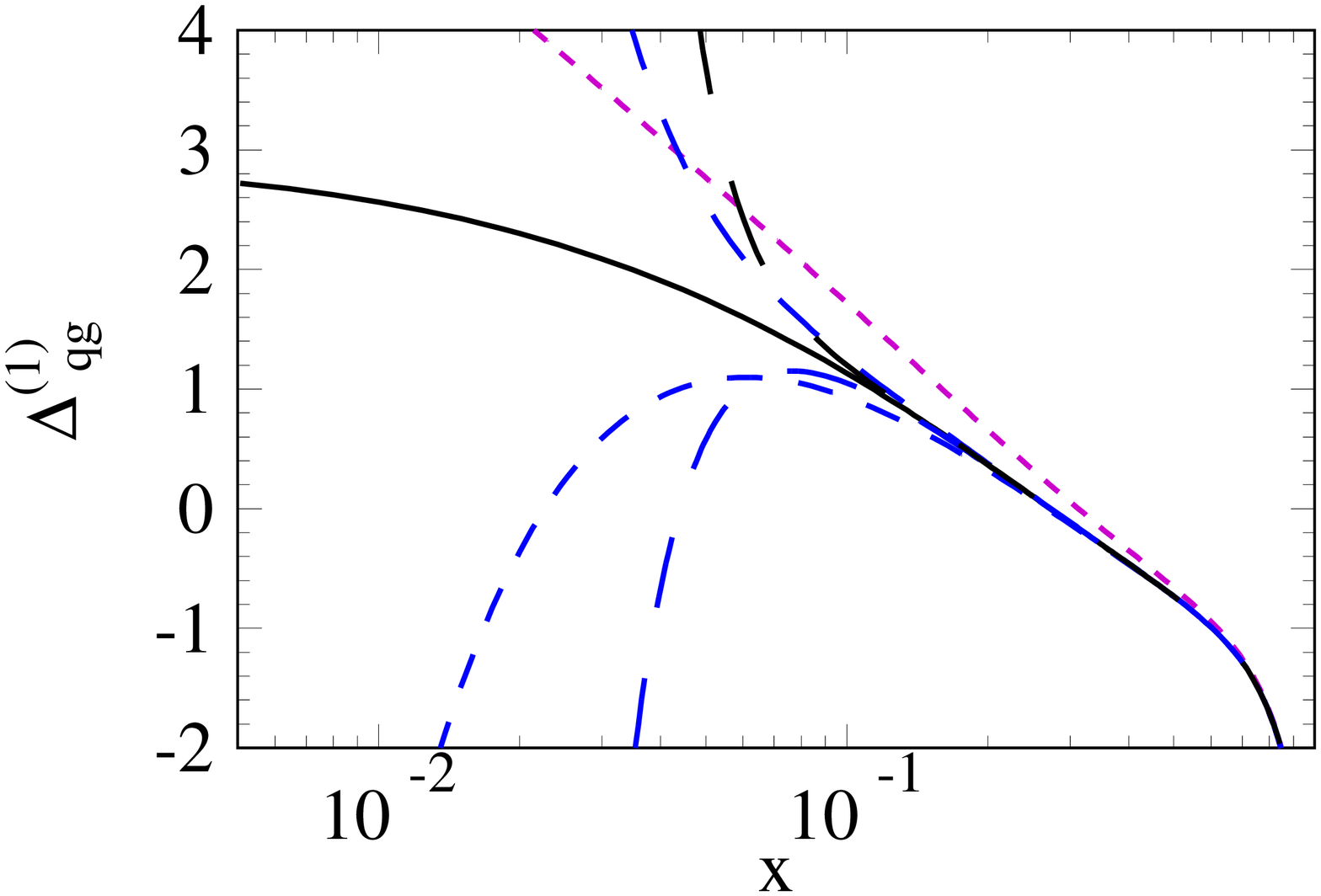}
    &
    \includegraphics[width=0.5\linewidth]{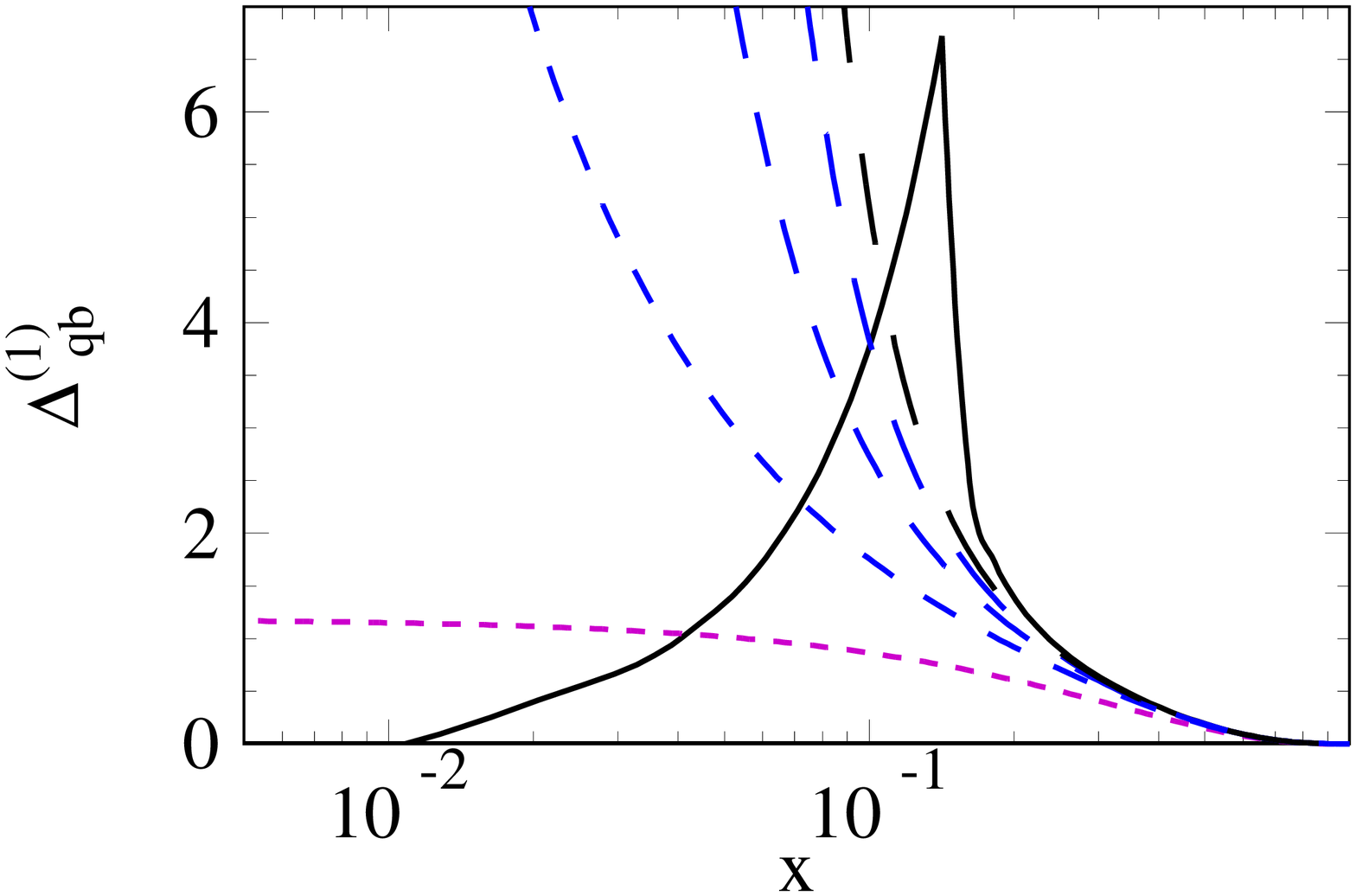}
    \\ (c) & (d)
  \end{tabular}
  \caption[]{\label{fig::NLOpart}Partonic NLO cross sections for the
    (a) $gg$, (c) $qg$ and (d) $q\bar{q}$ channel as functions of $x$ for
    $M_H=130$~GeV. The expansion
    in $\rho\to0$ (dashed lines) is compared with the exact result (solid
    lines). Lines with longer dashes include higher order terms in $\rho$.
    The interpolation (see text) is shown as a dotted line.
    In order to demonstrate the smallness of the region below
    threshold we show in (b) the result for the $gg$ channel using the
    linear $x$-scale. The vertical line at $x = M_H^2/(4M_t^2)\approx0.14$
    indicates the threshold for top quark pair production.}
\end{figure}

In Fig.~\ref{fig::NLOpart} functions $\Delta^{(1)}_{ij}$ in the three channels
are shown depending on $x$, evaluated for $M_H=130$~GeV and
$M_t=173.1$~GeV~\cite{:2009ec}, 
including successively higher orders in $\rho$ (dashed lines). The exact result
is plotted as a solid line. One sees that the leading term in $\rho$ is smooth
and demonstrates a reasonably good agreement with the exact
curve.\footnote{For $q\bar{q}$ the agreement is not that obvious, however, the
hadronic results have the proper order of magnitude.}
However, the
higher order terms in $\rho$ introduce divergences at $x\to 0$ which are the 
most severe for the $q\bar{q}$ channel.
This signifies the breakdown of the
assumption that $M_t^2\gg \hat{s}$ for large $\hat{s}$.
Note, however, the decent convergence above the threshold for the
top quark pair production, i.e., for $x > x_{th} = M_H^2/(4M_t^2)$.

In order to reduce the dependence on unphysical divergences near $x = 0$ 
at the NNLO, where exact cross sections are not known, we here devise two 
practical recipes and test them against the NLO results.

For the quark channels no information beyond $\rho$ expansion is available.
Thus, considering that these contributions are numerically suppressed and that 
$x_{th}$ limits the applicability of our asymptotic expansion,
we use the following
\begin{itemize}
\item[]Option~1:\\ For $x > x_{th}$, we use the complete result including all
  known $\mathcal{O}(\rho^n)$ corrections, and for $x < x_{th}$
  the infinite top mass approximation.
\end{itemize}

As will be demonstrated in the following section, this introduces an 
error in the hadronic results that is $<50\%$, which, if also true at the
NNLO, has a sufficiently small effect on the hadronic cross section compared
to the overall scale uncertainty.

For the primary production channel, $gg\to H$, NLO and NNLO asymptotics
near $x\to 0$ have been found in Ref.~\cite{Marzani:2008az}. Also, NLO 
plots suggest that in this channel there are no pronounced threshold 
effects at $x = x_{th}$. Thus, here we can use a more educated
\begin{itemize}
\item[]Option~2:\\ Complete $\mathcal{O}(\rho^n)$ result is matched 
at some point $x_m < x_{th}$ to a function $3 C_1  + ax$ (NLO) or 
$- 9 C_2 \ln{x} + b$ (NNLO), where coefficients $C_1$ and $C_2$ 
are tabulated in Ref.~\cite{Marzani:2008az} and $a$, $b$ and
the matching point $x_m$ are chosen to provide the most ``natural''
smooth behaviour of the function.
\end{itemize}

Note that this recipe is different from the procedures suggested 
in Ref.~\cite{Marzani:2008az} and Ref.~\cite{Harlander:2009mq}. The reason
is that our results (as in Eq.~(\ref{eqn:nlogg})) contain genuine
$1/x^n$ poles in $\mathcal{O}(\rho^n)$
contributions that are not present in the infinite top mass result
used in Ref.~\cite{Marzani:2008az} and are masked by the soft
expansion of Ref.~\cite{Harlander:2009mq}.

We found that an $x_m$ such that the function and its first derivative match 
smoothly is a good choice at the NLO; at the NNLO, matching at $x_m = x_{th}/4$ 
produces reasonable results for $110~{\rm GeV} \le M_H \le {\rm 300}$~GeV and
is consistent with the region of $x$ where higher $\mathcal{O}(\rho^n)$
corrections demonstrate good convergence. By varying the constants and interpolating
function shapes we have checked that the dependence of the hadronic
cross section on the exact details of the matching procedure is quite
small and that only the asymptotics near $x\to 0$ are important.

In Figs.~\ref{fig::NLOpart} (a) and (b) the thus obtained NLO extrapolation 
in the $gg$ channel is shown as a dotted line. One observes very good 
agreement with the exact curve, and the difference in the hadronic cross
sections is negligible. 


\subsection{NNLO corrections}

Combining the virtual part of $\Delta_{gg}^{(2)}$ calculated in
Refs.~\cite{Harlander:2009bw,Pak:2009bx} with the real contributions
we arrive at the $\mathcal{O}(\rho^n)$ corrections to the 
quantities $\Delta^{(2)}_{ij}$ for $n = 0,1,2$ for $gg$ and $qg$
reactions, and $n = 0,1,2,3$ for the remaining channels.
Our results expressed in terms of harmonic polylogarithms are quite
lengthy and can be obtained from the authors on request.
The $\mathcal{O}(\rho^0)$ terms exactly reproduce the 
expressions found in Ref.~\cite{Anastasiou:2002yz}. Expanding the
higher $\mathcal{O}(\rho^n)$ corrections in $(1-x)\ll 1$ we find
complete agreement with Ref.~\cite{Harlander:2009mq}.

\begin{figure}[t]
  \centering
  \begin{tabular}{cc}
    \includegraphics[width=0.46\linewidth]{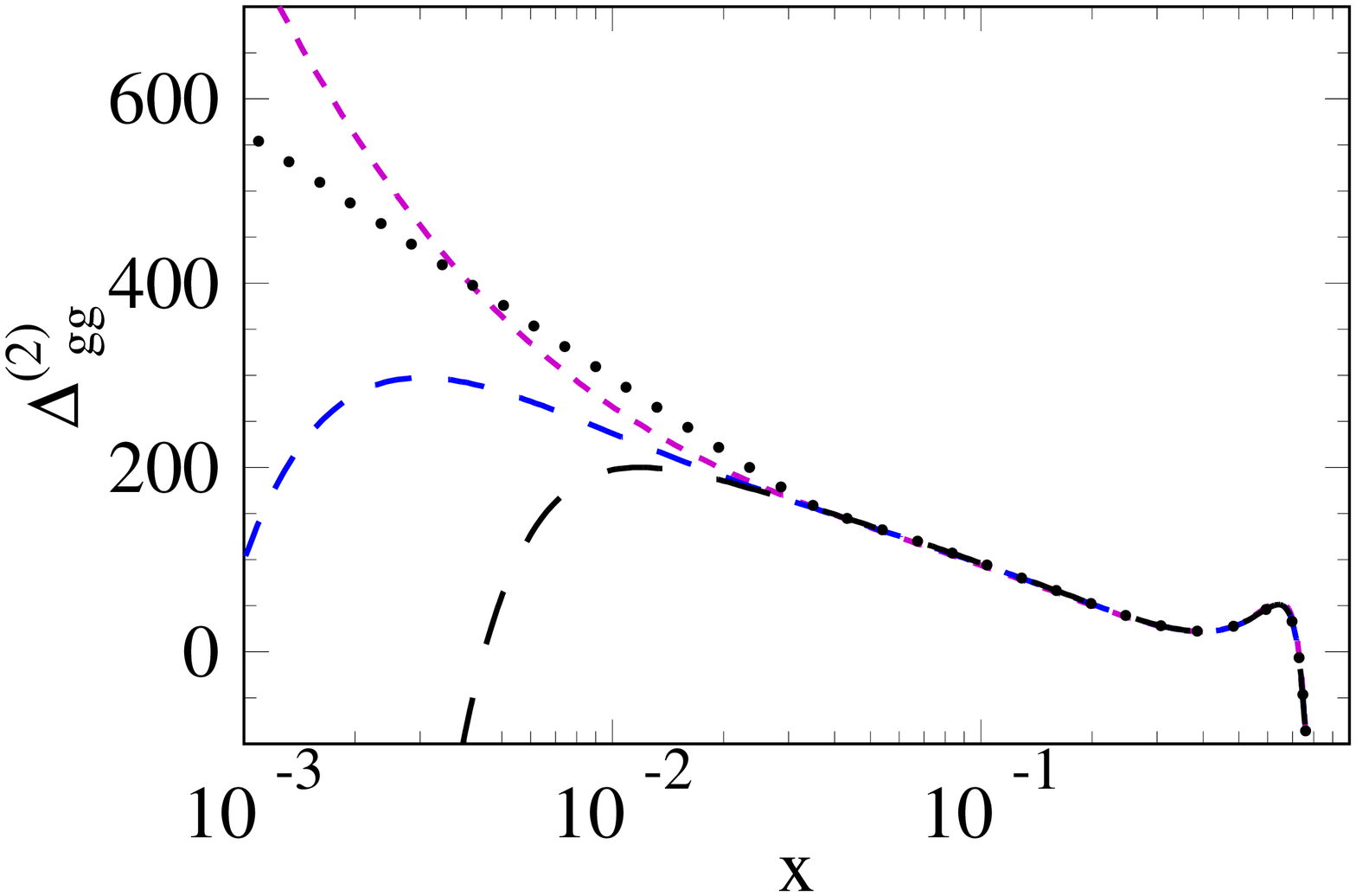}
    &
    \includegraphics[width=0.46\linewidth]{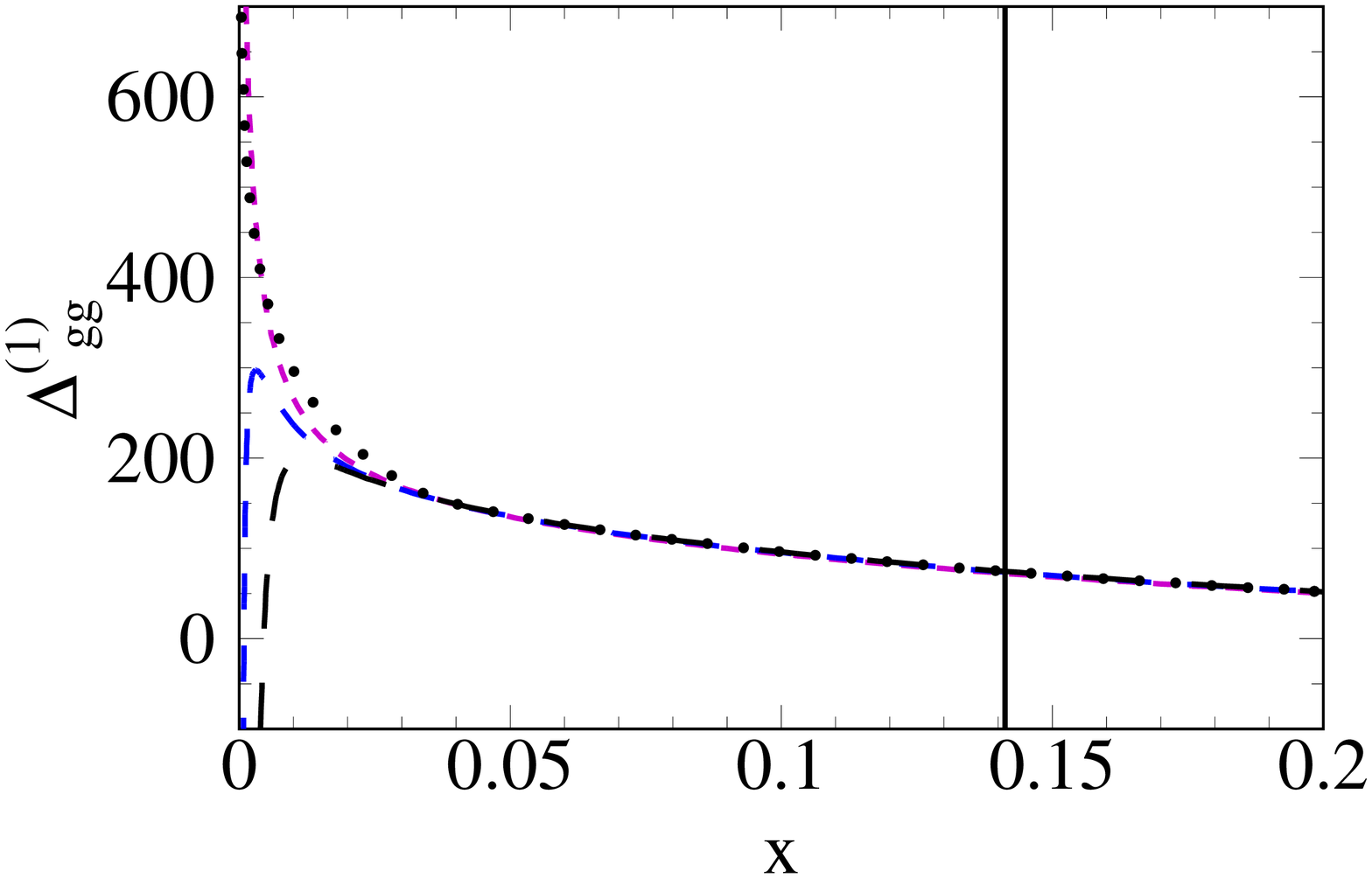}
    \\[-.5em] (a) & (b) \\[-1em]
    \includegraphics[width=0.46\linewidth]{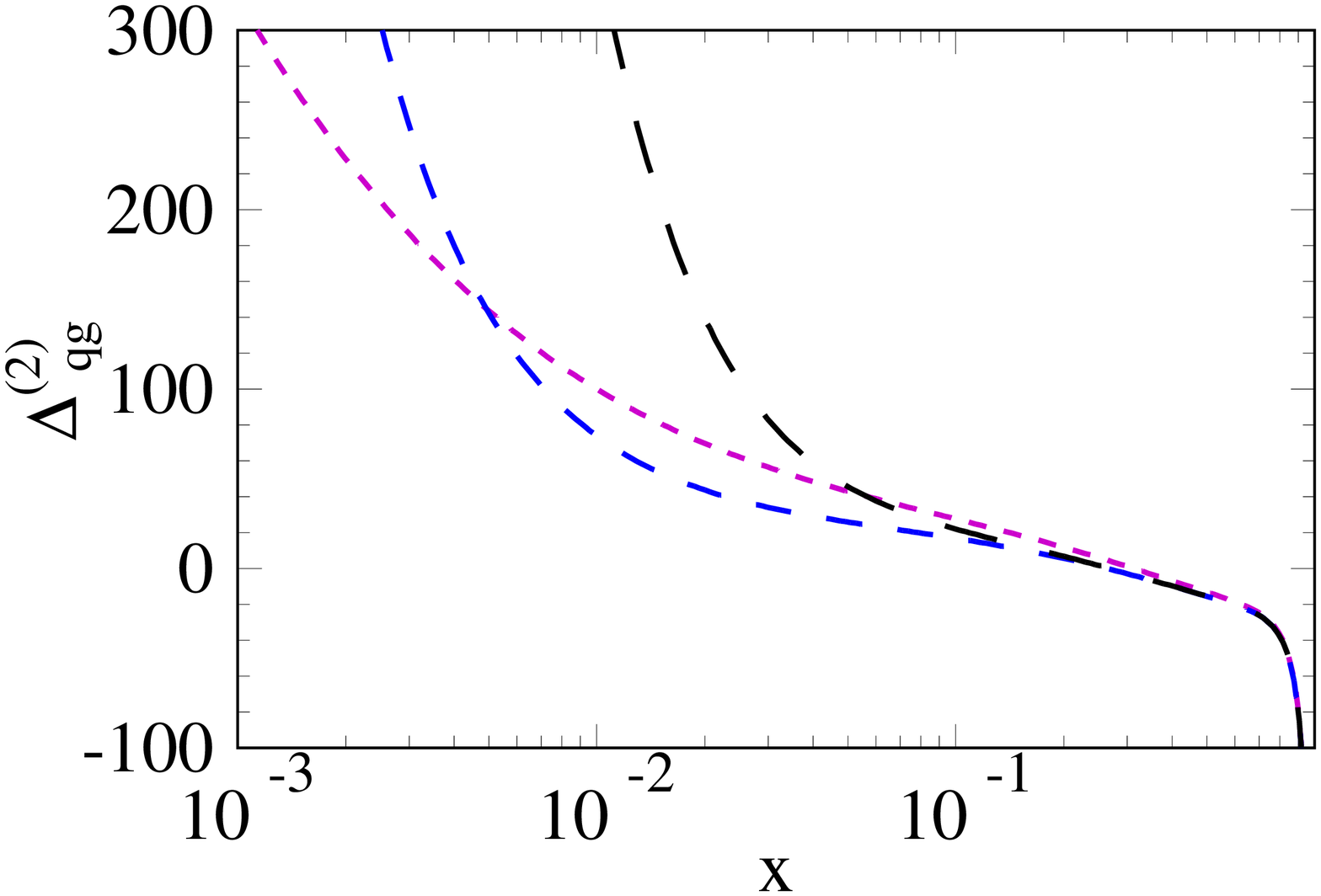}
    &
    \includegraphics[width=0.46\linewidth]{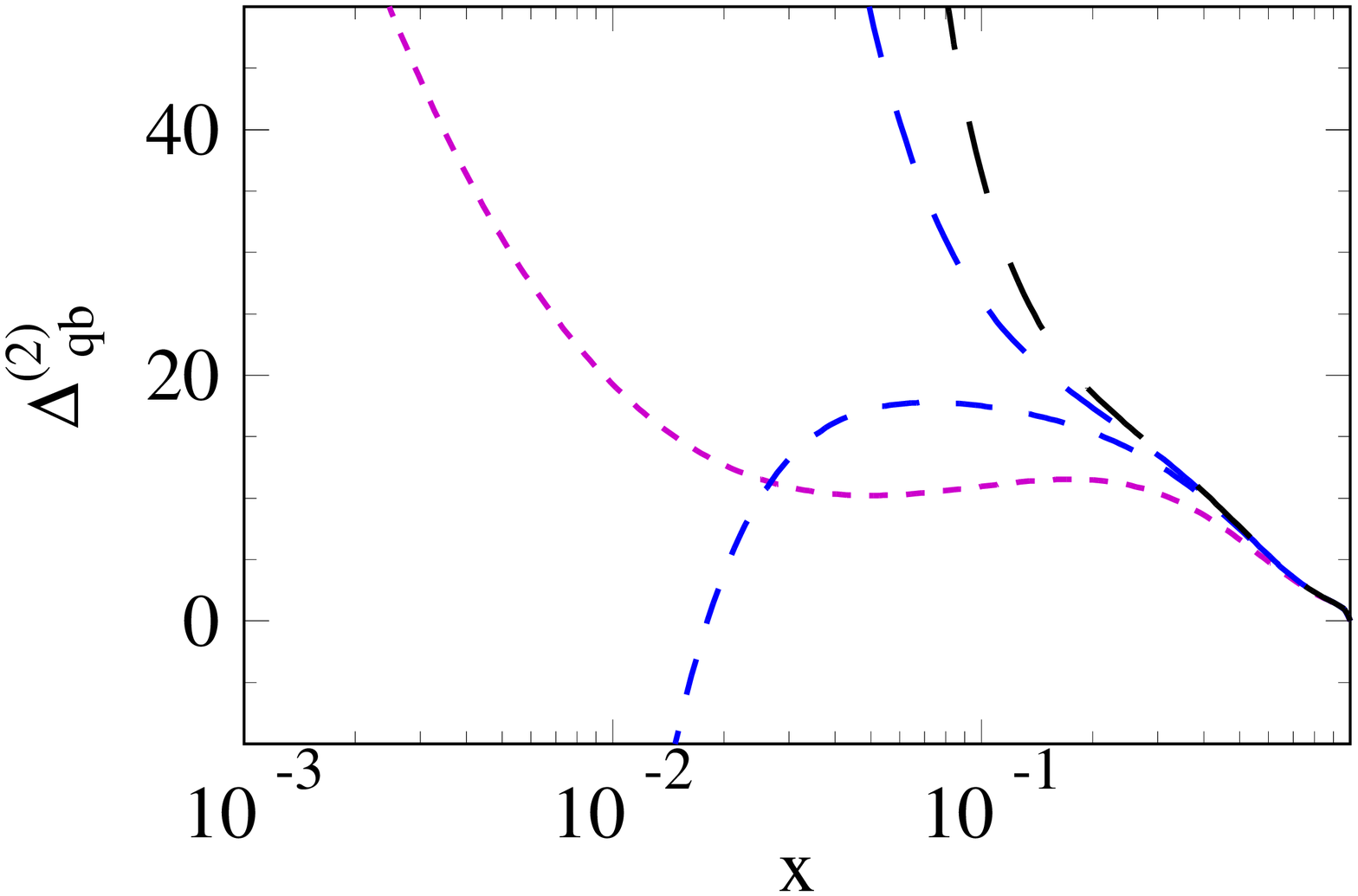}
    \\[-.5em] (c) & (d) \\[-1em]
    \includegraphics[width=0.46\linewidth]{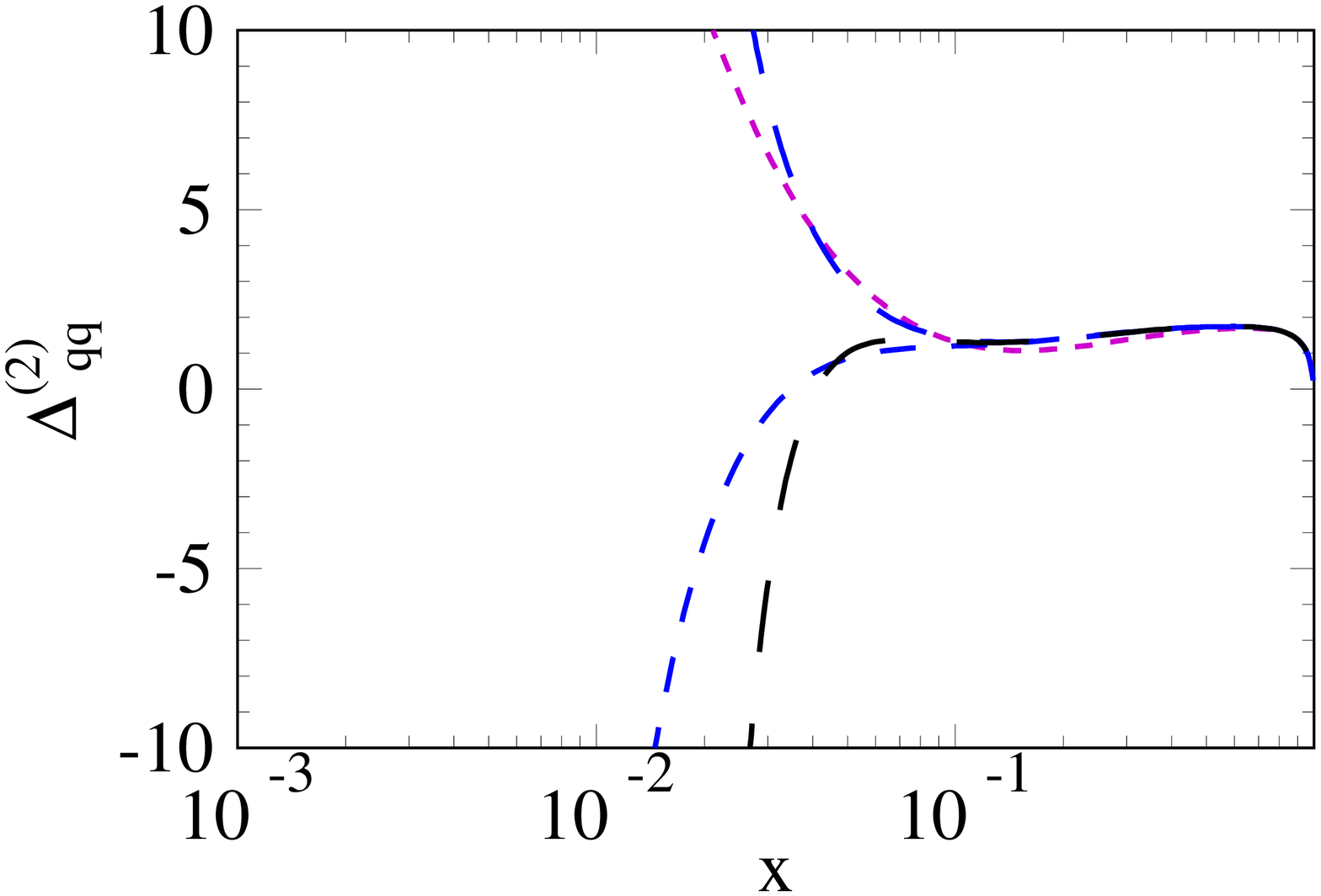}
    &
    \includegraphics[width=0.46\linewidth]{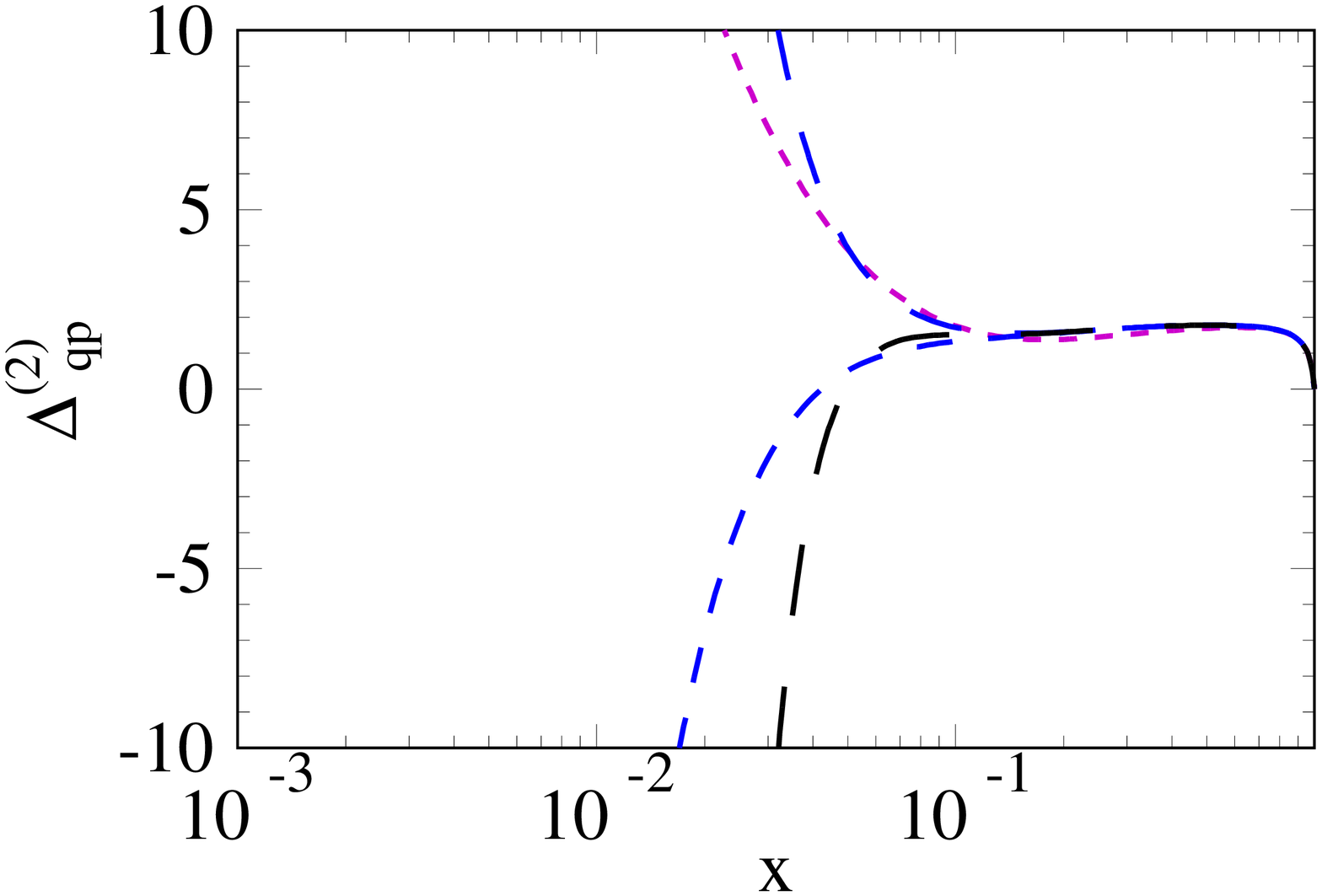}
    \\[-.5em] (e) & (f)
  \end{tabular}
  \caption[]{\label{fig::NNLOpart}Partonic NNLO cross sections for
    the (a) $gg$, (c) $qg$, (d) $q\bar{q}$, (e) $qq$, (f) $qq^\prime$ 
    channels functions of $x$ for $M_H=130$~GeV.
    Lines with longer dashes include higher order terms in $\rho$.
    In (b) we also show the $gg$ channel in the linear scale.
    The dotted line in (a) and (b) corresponds to the matched
    result.}
\end{figure}

In Fig.~\ref{fig::NNLOpart} we present non-singular parts of the 
functions $\Delta^{(2)}_{ij}$ for
${ij}={gg,qg,q\bar{q},qq,qq^\prime}$ as functions of $x$. Here one can
observe a behaviour similar to that at the NLO: the higher order
terms in $\rho$ develop more severe singularities near $x\to0$, 
however, below the threshold the results converge. The dotted curves
in Figs.~\ref{fig::NNLOpart}(a) and (b) demonstrate the ``Option 2''
extrapolations described above. The further numerical analysis is 
based on these extrapolations.

\section{\label{sec::hadr}Hadronic cross section}

The hadronic cross sections are given by the convolution of
the partonic cross section $\hat{\sigma}_{ij\to H+X}$ 
with the corresponding parton distribution functions (PDFs), which is
conventionally written as follows:\footnote{In this paper we concentrate
on $pp$ collisions at the LHC peak energy $\sqrt{s} = 14~{\rm TeV}$.
The modifications for $p\bar{p}$ collisions at the Tevatron are obvious.}

\begin{eqnarray}
  \sigma_{pp^\prime\to H+X}(s)
  &=&
  \sum_{k,l\in\{g,u,...,b,\bar{u},...,\bar{b}\}}
  \int_{M_H^2/s}^1 {\rm d}\tau\,
  \bigg[\frac{{\rm d}{\cal L}_{kl}}{{\rm d}\tau}\bigg](\tau,\mufs)~
  \hat{\sigma}_{kl\to H+X}(\hat s=\tau s,\mufs)\,,
  \label{eq::sighad0}
\end{eqnarray}
where ${\rm d}{\cal L}_{kl}/{\rm d}\tau$ is the so-called luminosity
function given by
\begin{eqnarray}
  \bigg[\frac{d{\cal L}_{kl}}{d\tau}\bigg](\tau,\mufs)
  &=&
  \int_0^1 {\rm d}x_1 \int_0^1 {\rm d}x_2\,
  f_{k/p}(x_1,\mufs) f_{l/p}(x_2, \mufs)\,
  \delta(\tau - x_1 x_2)\,.
  \label{eq::lumi0}
\end{eqnarray}

\begin{figure}[t]
  \centering
  \includegraphics[width=\linewidth]{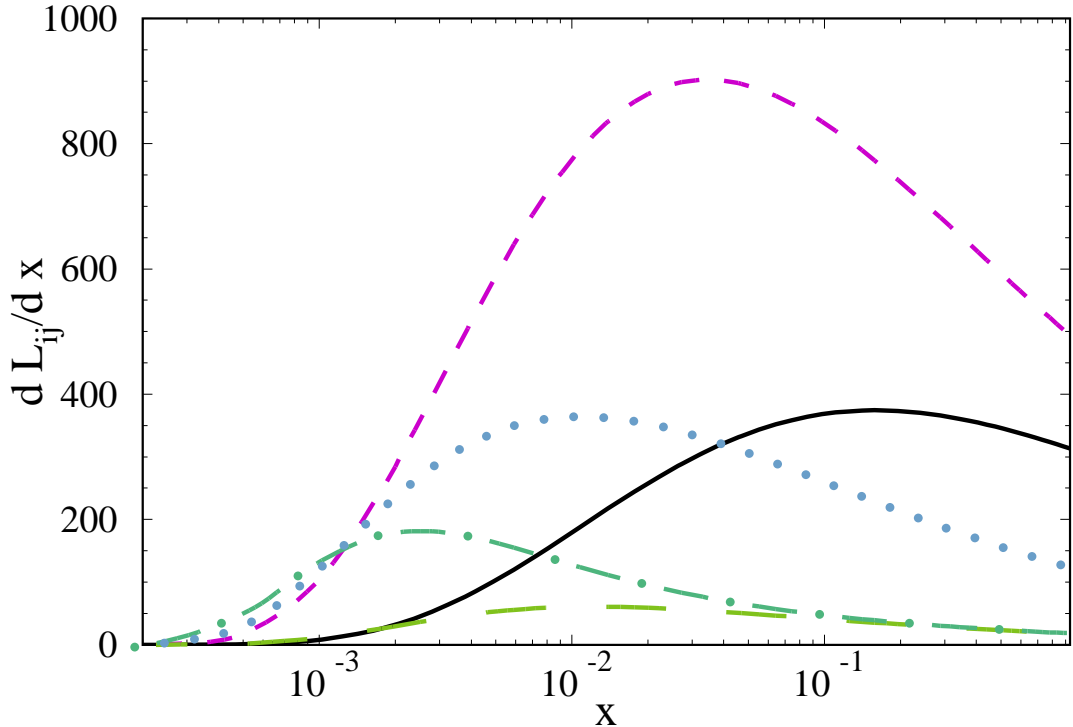}
  \caption[]{\label{fig::lumf}Luminosity functions for $ij=gg$ (solid), $qg$
    (short dashed), $q\bar{q}$ (long dashed), $qq$ (dash-dotted), and
    $qq^\prime$ (dotted line).}
\end{figure}

For the further discussion we adopt a slightly different
parametrization in terms of the ``natural'' parameter $x = M^2_H/\hat{s}$
and the distinct production channels:
\begin{eqnarray}
  \sigma_{pp^\prime\to H+X}(s)
  &=&
  \sum_{ij\in\{gg,~ qg,~ q\bar{q},~ qq,~ qq^\prime\}}
  \int_{M_H^2/s}^1 {\rm d}x\,
  \bigg[\frac{{\rm d}{\cal L}_{ij}}{{\rm d}x}\bigg](x,\mufs)~
  \hat{\sigma}_{ij\to H+X}(x,\mufs)\,,
  \label{eq::sighad}
\end{eqnarray}
with the straightforward modifications to the corresponding weights.
For example, the quark-gluon luminosity is defined as
\begin{eqnarray}
  \bigg[\frac{d{\cal L}_{qg}}{{\rm d}x}\bigg](x,\mufs)
  &=&
  2 \sum_{q\in\{u,...,b,\bar{u},...,\bar{b}\}}
  \int_0^1 {\rm d}x_1 \int_0^1 {\rm d}x_2\,
  f_{g/p}(x_1,\mufs) f_{q/p}(x_2,\mufs)\,
  \\ \nonumber && \times
  \delta\left({M^2_H\over s x} - x_1 x_2\right) {M^2_H \over s x^2}\,. 
  \label{eq::lumi} 
\end{eqnarray}
For illustration we show in Fig.~\ref{fig::lumf} the quantities
${\rm d}{\cal L}_{ij}/{\rm d}x$ for $ij=gg$, $qg$, $q\bar{q}$,
$qq$, and $qq^\prime$.
One notices that for $x\to 0$ there is a rapid decay of all luminosity
functions which is one of the main reasons that at the NLO the heavy top
approximation works extremely well.

We use the parton distribution function (PDF) set MSTW2008~\cite{Martin:2009iq}
and the $\alpha_s$ evolution at LO, NLO and NNLO when computing predictions
to the cross section at the corresponding order.
To discuss the numerical effect of our calculation we decompose the prediction
of the total cross section into its LO, NLO and NNLO contributions:
\begin{eqnarray}
  \sigma_{pp^\prime\to H+X}(s) &=&   
  \sigma^{\rm LO} + \delta\sigma^{\rm NLO} + \delta\sigma^{\rm NNLO}
  \,,
  \label{eq::deltasigma}
\end{eqnarray}
and denote the heavy top quark approximation with an additional
subscript $\infty$.

\begin{figure}[t]
  \centering
  \begin{tabular}{cc}
    \includegraphics[width=0.5\linewidth]{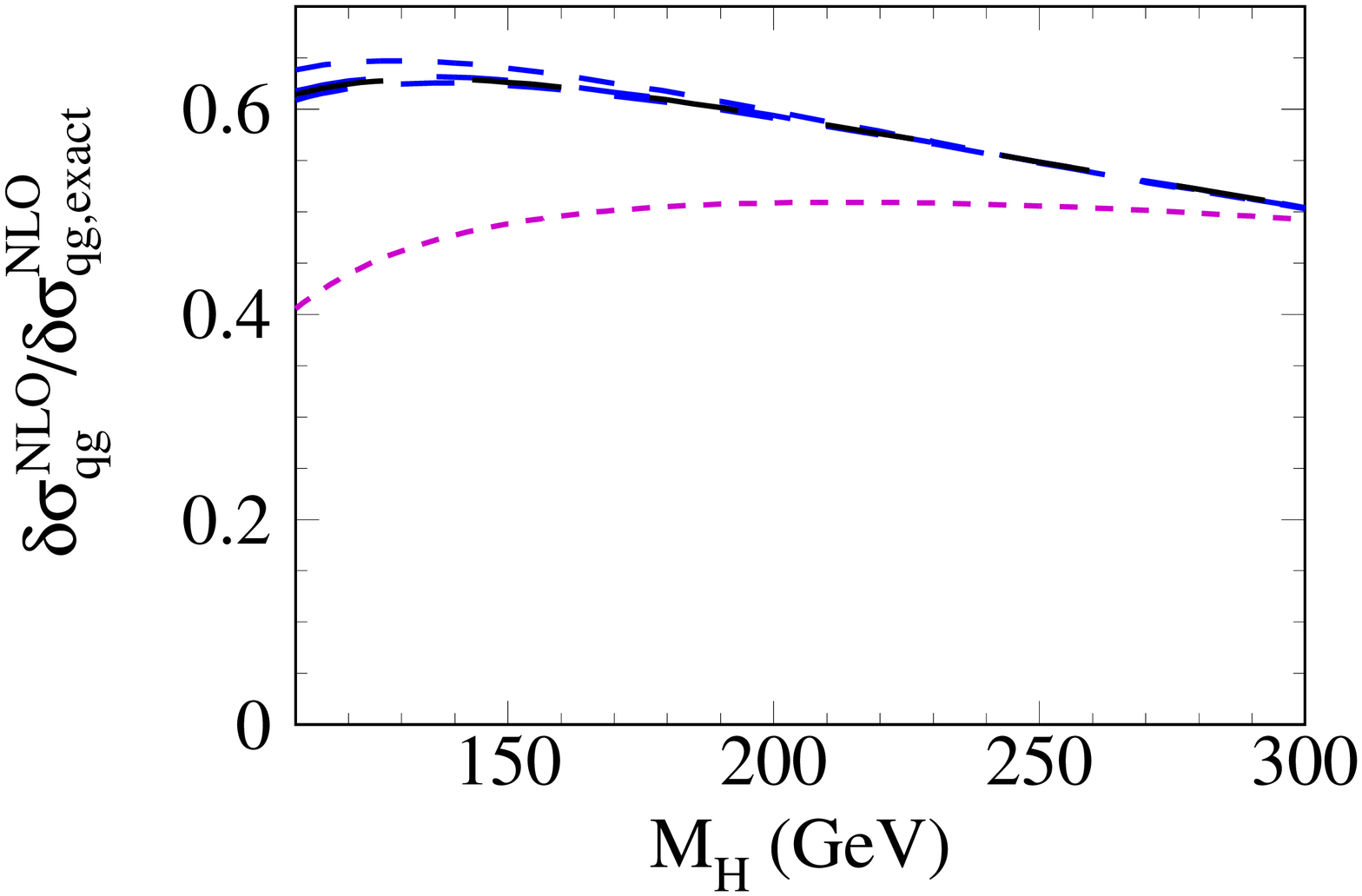}
    &
    \includegraphics[width=0.5\linewidth]{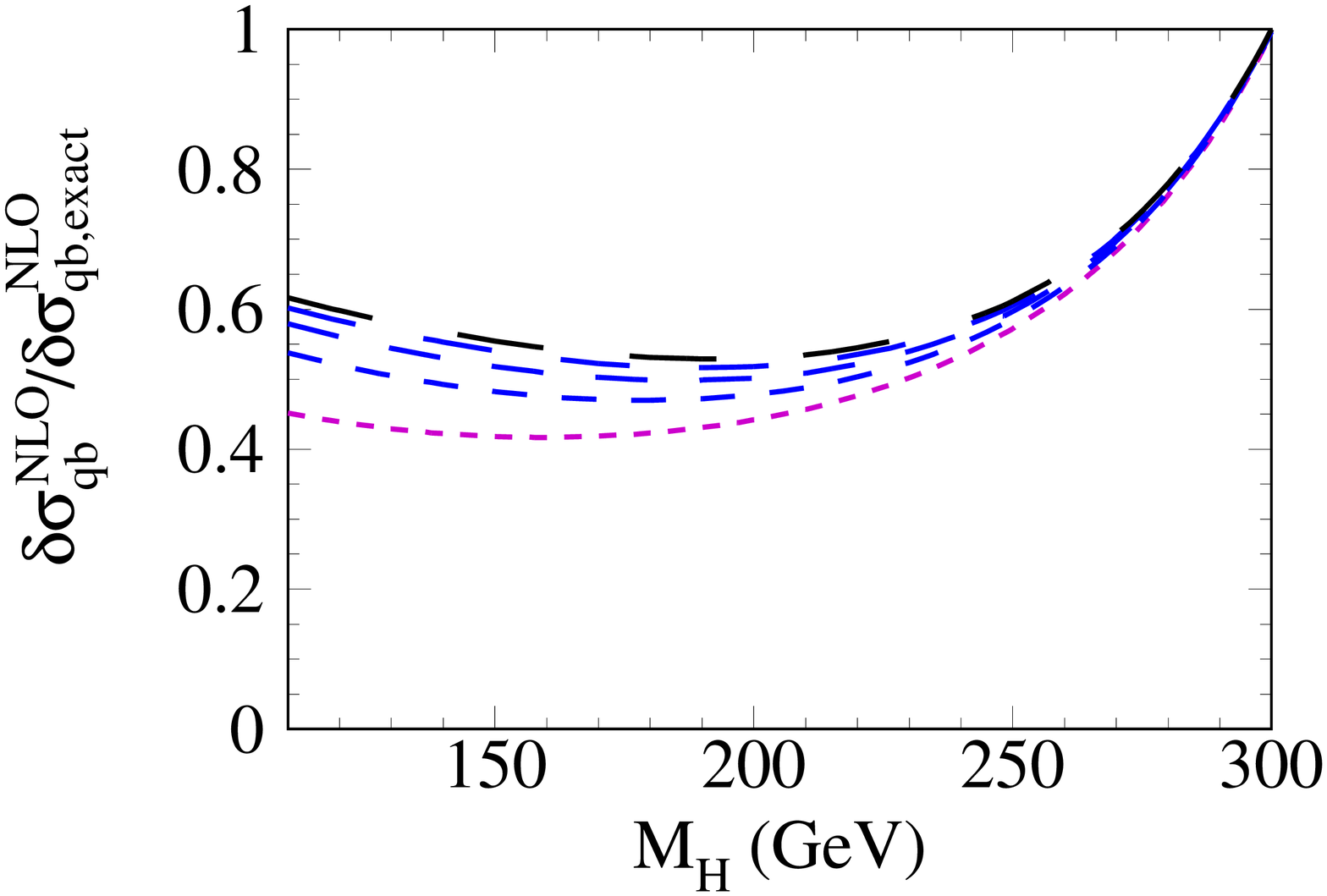}
    \\ (a) & (b) \\
  \end{tabular}
  \caption[]{\label{fig::opt2_nlo}Ratio of the NLO hadronic cross section 
    including successive higher orders in $1/M_t$ (from short to long dashes)
    normalized to the exact result. (a) $qg$ and (b) $q\bar{q}$ }
\end{figure}

Let us in a first step discuss the channels involving quarks which are
treated using ``Option~1'' as described in Section~\ref{sec::part}.
In Fig.~\ref{fig::opt2_nlo} we show the $M_H$-dependence of the 
NLO contribution to the hadronic cross section originating from the 
quantity $\Delta_{ij}^{(1)}$ (cf. Eq.~(\ref{eq::hatsigma}))
normalized to the exact result 
as coded in HIGLU~\cite{Spira:1995mt} for $M_H$ between 110~GeV and 300~GeV.
For the $qg$ channel one observes that the infinite top quark mass
approximation provides between 40 and 50\% of the exact result. After
including the $\rho$ and $\rho^2$ term this is improved to about 60\% for the
smaller Higgs boson masses whereas for $M_H=300$~GeV the heavy top quark
mass is practically unchanged.
Similarly, for the $q\bar{q}$ channel we observe an improvement 
by about 10 to 15\% for the Higgs boson masses around 140~GeV.
For $M_H=300~{\rm GeV}$ the heavy top expansion is practically equivalent to
the exact result.

\begin{figure}[t]
  \centering
  \begin{tabular}{cc}
    \includegraphics[width=0.5\linewidth]{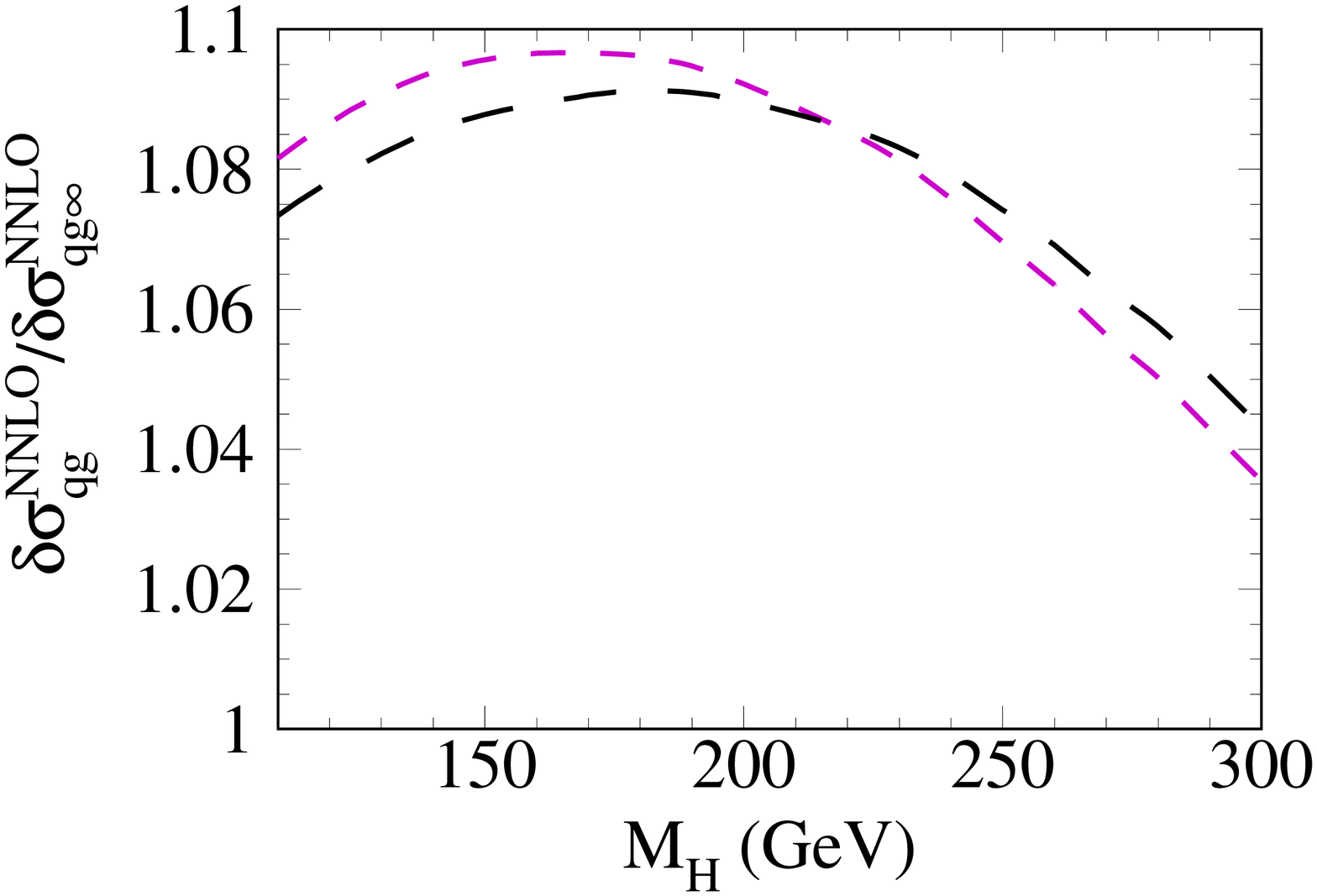}
    &
    \includegraphics[width=0.5\linewidth]{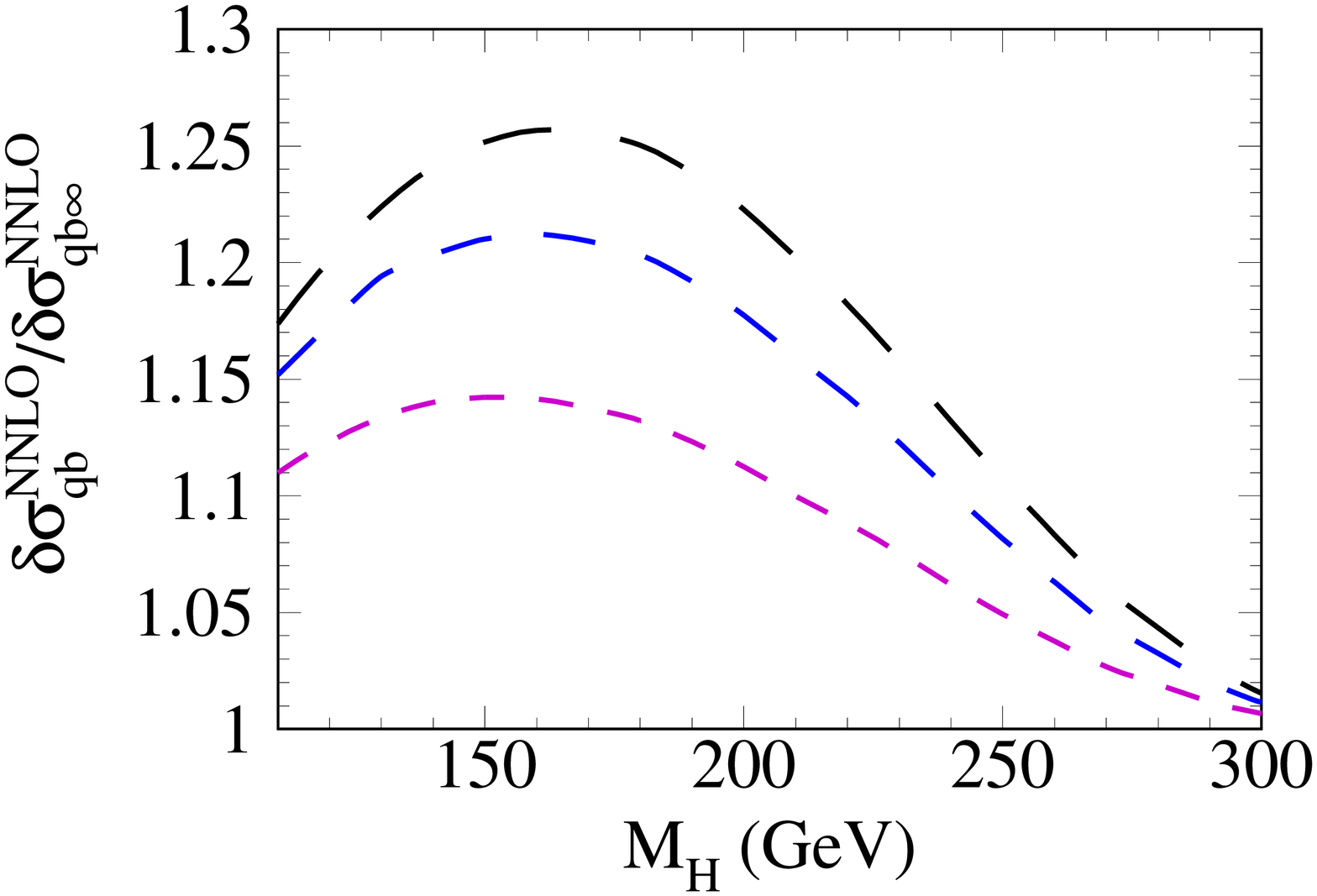}
    \\ (a) & (b) \\
    \includegraphics[width=0.5\linewidth]{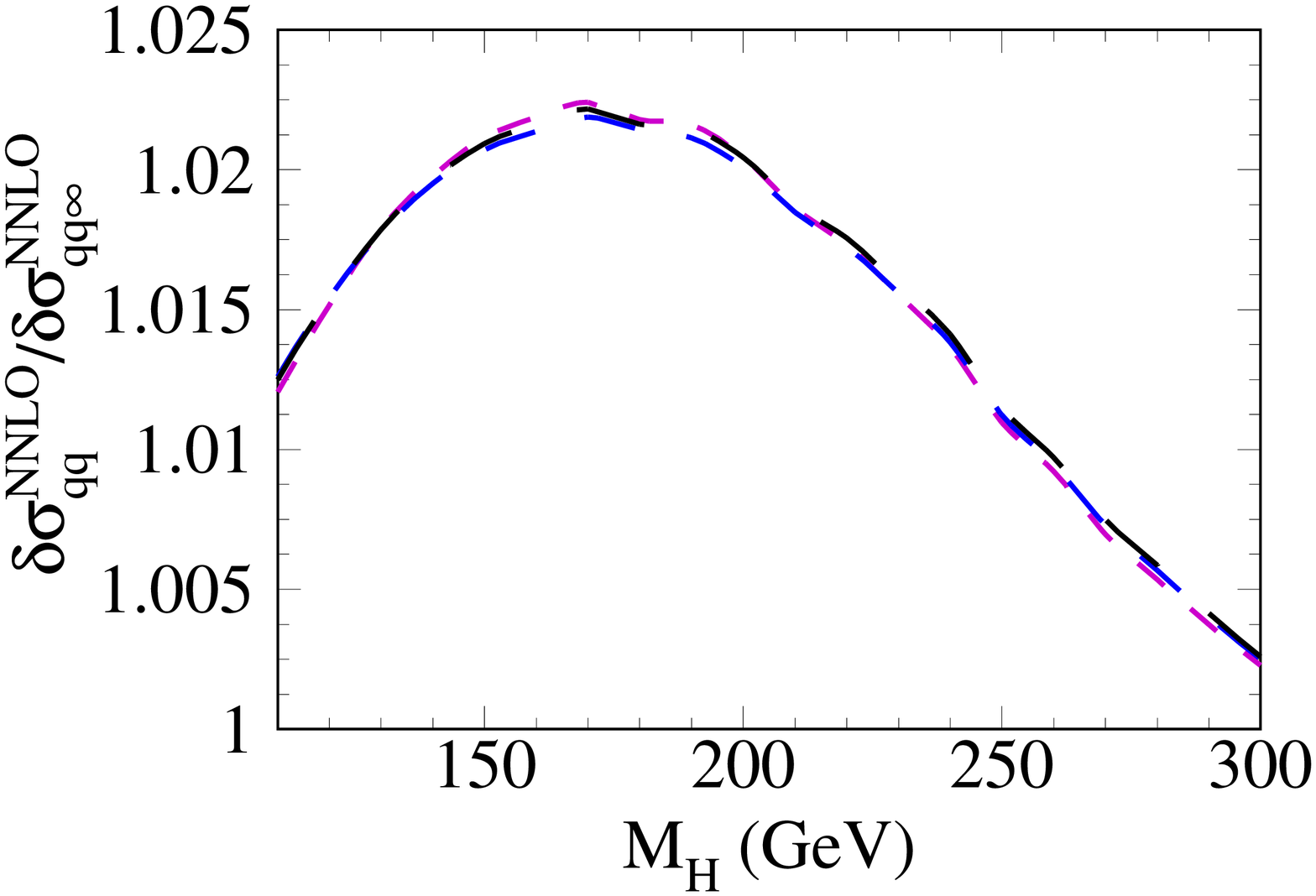}
    &
    \includegraphics[width=0.5\linewidth]{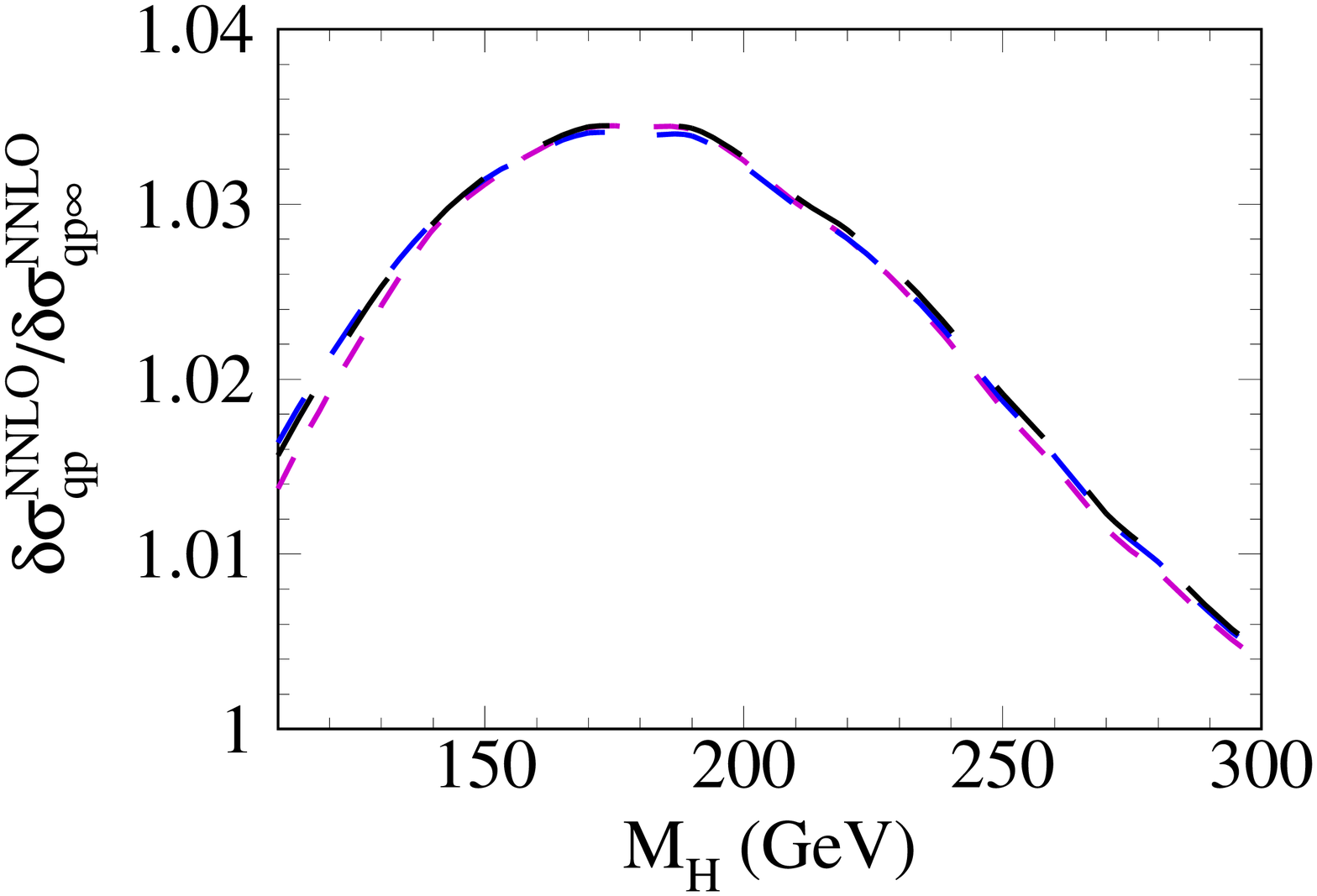}
    \\ (c) & (d) \\
  \end{tabular}
  \caption[]{\label{fig::opt2_nnlo}Ratio of the NNLO hadronic cross section 
    including successive higher orders in $1/M_t$ (from short to long dashes)
    normalized to the infinite top quark mass result.
    (a) $qg$, (b) $q\bar{q}$, (c) $qq$, (d) $qq^\prime$.}
\end{figure}

The analogous curves at NNLO can be found in
Fig.~\ref{fig::opt2_nnlo} where we normalize the result on the
infinite top quark mass approximation. In all cases the power-suppressed terms
lead to an increase of the cross section between 4\% and
10\% for the quark-gluon and up to 25\% for the quark-anti-quark channel
in our range of Higgs boson masses.
The very rapid convergence is observed for the $qq$ and $qq^\prime$
channels where the contribution beyond the $1/M_t^2$ term is practically zero.

\begin{figure}[t]
  \centering
  \begin{tabular}{cc}
    \includegraphics[width=0.5\linewidth]{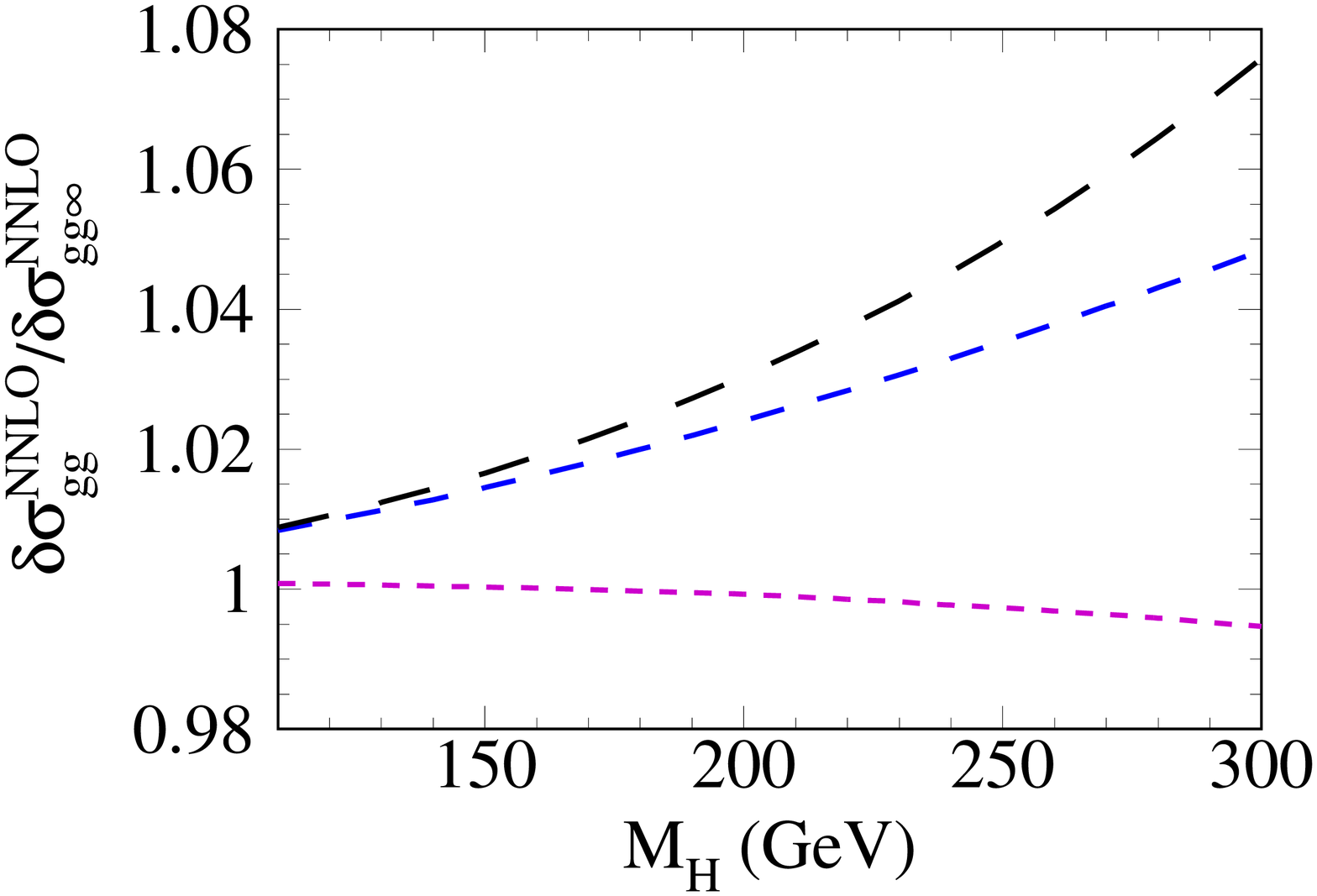}
    &
    \includegraphics[width=0.5\linewidth]{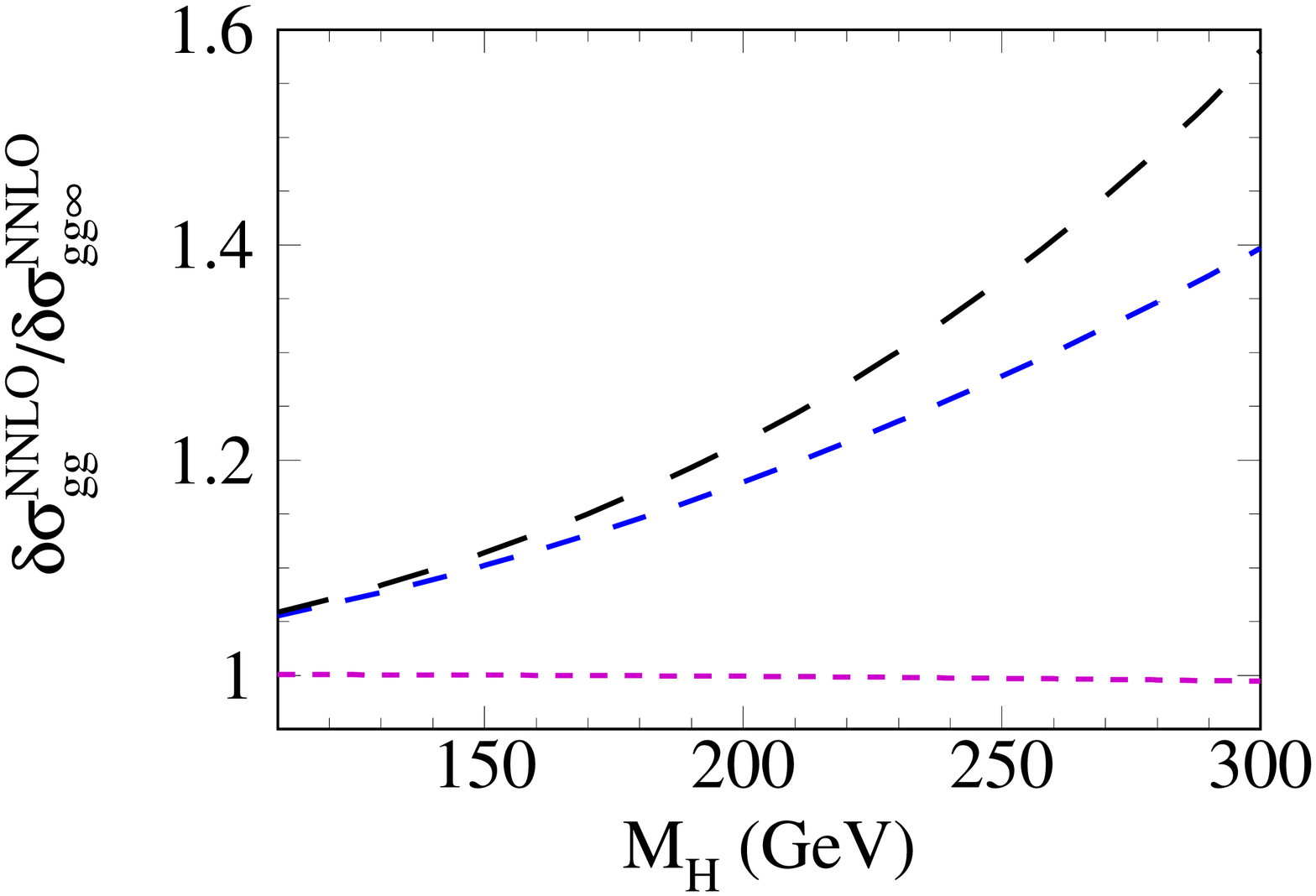}
    \\ (a) & (b) \\
    \includegraphics[width=0.5\linewidth]{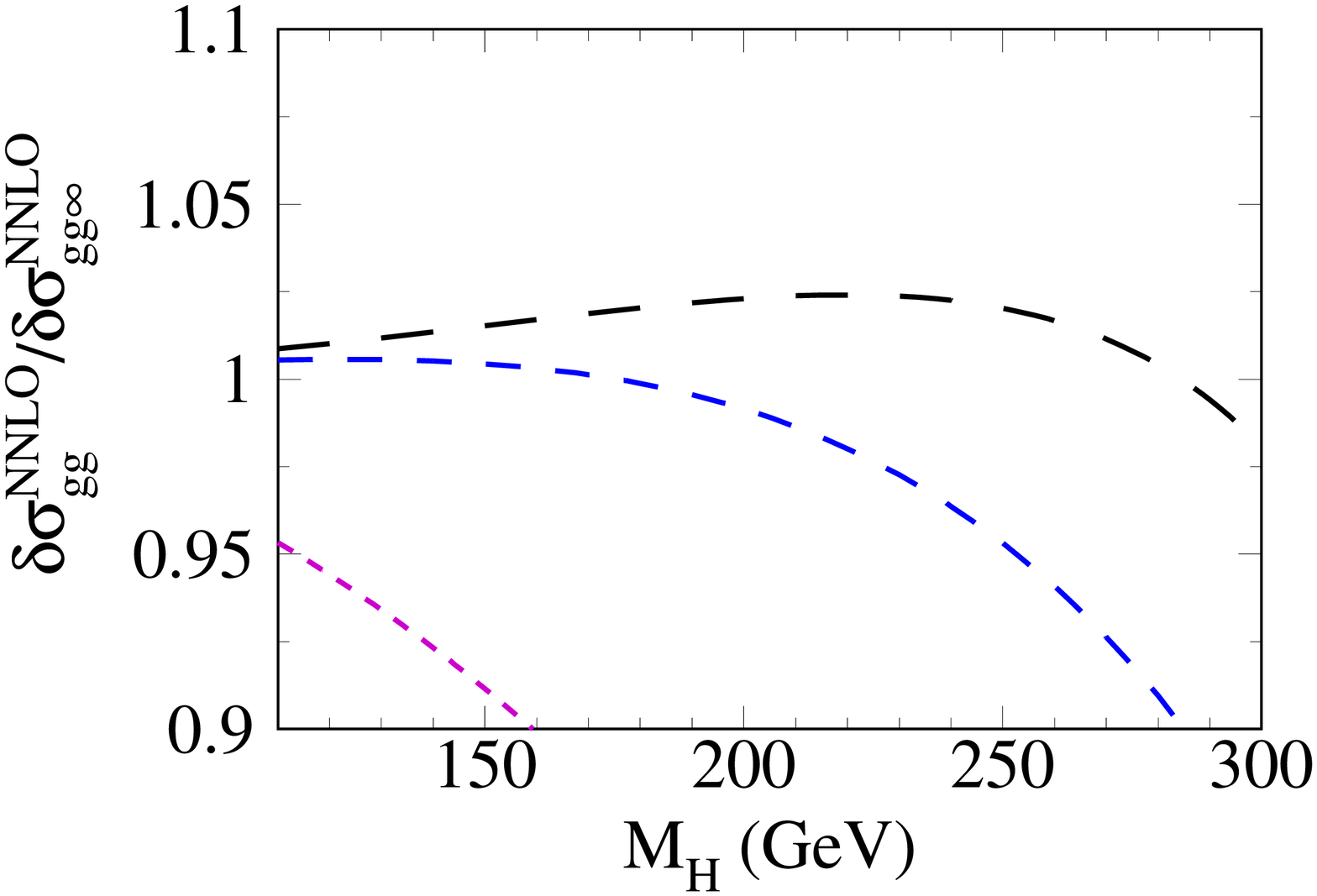}
    &
    \includegraphics[width=0.5\linewidth]{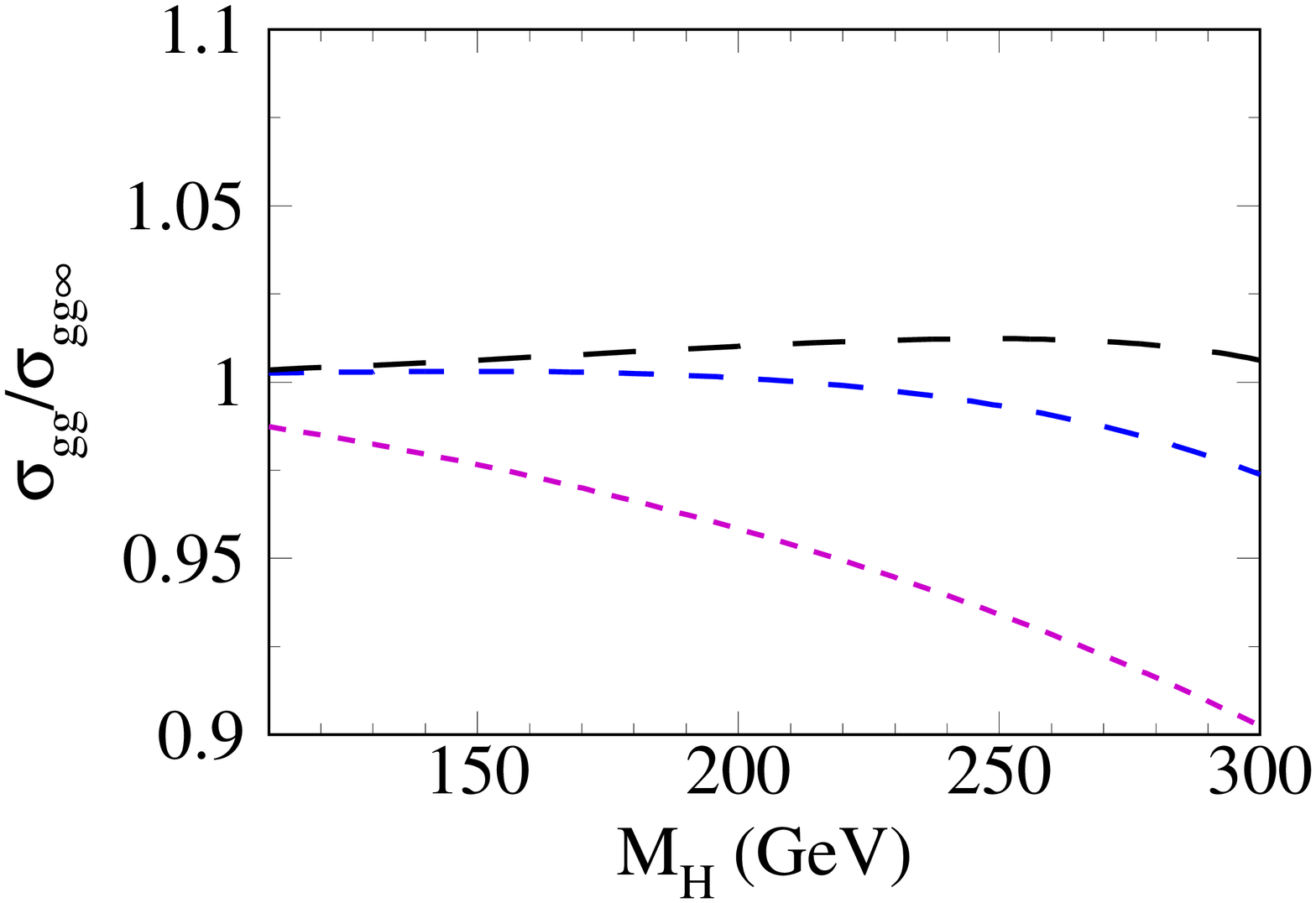}
    \\ (c) & (d) \\
  \end{tabular}
  \caption[]{\label{fig::opt1_nnlo}(a), (b) and (c):
    Ratio of the NNLO hadronic cross section 
    ($gg$ contribution) including successive higher orders in $1/M_t$ normalized
    to the infinite top quark mass result. In (a) the exact LO mass
    dependence is factorized both in the numerator and denominator.
    In (b) numerator and denominator are expanded in $\rho$, and 
    in (c) only the numerator is expanded.
    (d) shows the prediction of the gluon-induced inclusive Higgs production
    cross section up to NNLO normalized to the heavy top limit.}
\end{figure}

Let us finally turn to the numerically most important contribution, the
$gg$ channel treated with the matching to $\hat{s}\to\infty$ asymptotics
denoted above as the ``Option~2''. In Figs.~\ref{fig::opt1_nnlo}(a)--(c) we
demonstrate the NNLO contribution to the hadronic cross section
(cf. Eq.~(\ref{eq::deltasigma})) normalized to the infinite top quark
mass result. The difference is that in~(a), the exact LO top quark mass dependence
is factored out as in Eq.~(\ref{eq::hatsigma}), while in~(b) the partonic
cross sections both in numerator and denominator are strictly expanded in $\rho$.
Finally, in~(c) we expand $\hat{A}_{LO}$ in the numerator
but keep it exact in the denominator.

For the fully expanded option~(b) one observes for $M_H=300$~GeV corrections
up to 40\% originating from the linear $\rho$ term which further increase to
almost 60\% after including the $\rho^2$ term.
However, when the exact leading-order top quark mass dependence is factored
out (case~(a)), the corrections amount to at most 8\%.
Considering the fact that the NNLO terms contribute about 10\% 
of the total NNLO cross section we conclude that the 
top quark mass suppressed terms at NNLO alter the prediction by less than
1\%. This justifies the use of the heavy top mass approximation for the
evaluation of the NNLO hadronic cross section.
The latter conclusion is also obtained from Fig.~\ref{fig::opt1_nnlo}(c).

In Fig.~\ref{fig::opt1_nnlo}(d) we also take into account the exact LO and NLO
contribution and again study the effect of the different $\rho$ terms. Similar
to the case~(c), we leave in the denominator the exact LO mass dependence and
consider various expansion depths of the NNLO contribution in the numerator.
This plot can be directly compared to the left panel of Fig.~7 in
Ref.~\cite{Harlander:2009mq}. Very good agreement is observed; the minor
differences can be traced back to the different matching procedures.

\section{\label{sec::concl}Conclusions}

In this paper we present the NNLO production cross section of the
Standard Model Higgs boson including finite top quark mass effects.
Our calculation is based on the evaluation of the imaginary part of the
forward scattering amplitudes which, via the optical theorem, directly
leads to the total cross section. We apply the asymptotic expansion in
order to obtain correction terms suppressed by the heavy top quark mass.

We observe rapid convergence of the series below the
threshold for the production of real top quarks, i.e. for $\hat{s}\le 4 M_t^2$.
However, the region of small $x=M_H^2/\hat{s}$ demonstrates $1/x^n$ singularities 
as a consequence of our expansion procedure. For the numerically dominant
gluon-gluon channel we match our results to the large $\hat{s}$ limit, 
curing thus those artificial singularities and obtaining stable predictions
for the hadronic cross section.

The numerical impact of the top quark mass suppressed terms is 
below approximately 1\% and thus about a factor ten smaller than the uncertainty
from scale variation. Let us, however, stress that this result was not 
obvious a priori. Our calculation justifies the use of the heavy top
quark mass approximation when evaluating the NNLO cross section.

In addition, we confirm the results of Ref.~\cite{Anastasiou:2002yz}
for the infinite top quark mass, and the $M_t$-suppressed terms calculated
in Ref.~\cite{Harlander:2009mq}.


\vspace*{2em}
{\large\bf Acknowledgements}

We thank Robert Harlander and Kemal Ozeren for providing us with their
analytical results and Kirill Melnikov for the useful communication.
This work was supported by the DFG through the SFB/TR~9 ``Computational
Particle Physics'' and by the BMBF through Grant No. 05H09VKE. M.R. was
supported by the Helmholtz Alliance ``Physics at the Terascale''. 




\end{document}